\documentclass[useAMS,usenatbib]{mn2e}

\usepackage{times}
\usepackage{graphics,epsfig}
\usepackage{graphicx}
\usepackage{amssymb}

\newcommand{\sgas}{$\Sigma_{\rm gas}$}
\newcommand{\shi}{$\Sigma_{\rm HI}$}
\newcommand{\ssfr}{$\Sigma_{\rm SFR}$}

\title[Star formation in extreme dwarfs]{Star formation in extremely faint dwarf galaxies}
\author[Roychowdhury et al.]{Sambit Roychowdhury$^{1}$\thanks{E-mail: sambit@ncra.tifr.res.in}, Jayaram N. Chengalur$^{1}$\thanks{~~~~~~~~~~~~~chengalu@ncra.tifr.res.in}, Ayesha Begum$^{2}$\thanks{~~~~~~~~~~~~~begum@astro.wisc.edu}
\newauthor and Igor D. Karachentsev$^{3}$\thanks{~~~~~~~~~~~~~ikar@sao.ru}\\
       \\ 
       $^{1}$NCRA-TIFR, Post Bag 3, Ganeshkhind, Pune 411 007, India\\
       $^{2}$Dept of Astronomy, University of Wisconsin-Madison, Madison WI 53706-1582\\
       $^{3}$Special Astrophysical Observatory, Russian Academy of Sciences, N. Arkhyz, KChR 369167, Russia}

\begin{document}
\date{}

\pagerange{\pageref{firstpage}--\pageref{lastpage}} \pubyear{}

\maketitle

\label{firstpage}

\begin{abstract}
We study the relationship between the gas column density (\sgas) and the star formation rate surface density (\ssfr) for a sample of 23 extremely faint dwarf irregular galaxies drawn from the Faint Irregular Galaxy GMRT Survey (FIGGS).  Our sample galaxies have a median HI mass of $2.8 \times 10^7 {\rm M}_\odot$ and a median blue magnitude M$_{\rm B} \sim -13.2$. \ssfr\  is derived from \emph{GALEX} data, while \sgas\ is derived from the GMRT based FIGGS HI 21 cm survey data. We find that \sgas\ averaged over the star forming region of the disk lies below most estimates of the ``threshold density'' for star formation, and that the average \ssfr\  is also lower than would be expected from the ``Kennicutt--Schmidt'' law \citep{ken98}. This deviation is indicative of an environmental dependence of star formation efficiencies, since the \cite{ken98} relation was derived from data on the central regions of large spiral and star burst galaxies. The star formation efficiency in small galaxies may be more relevant to modeling of star formation in gas rich, low metallicity systems in the early universe. We also use our data to look for small scale (400 pc and 200 pc) correlations between \ssfr\ and \sgas. For 18 of our 23 galaxies, we find that \ssfr\ can be parametrized as having a power law dependence on \sgas. The power law relation holds until one reaches the sensitivity limit of the \emph{GALEX} data, i.e. we find  no evidence for a ``threshold density'' below which star formation is completely cut off. The power law slopes and coefficients however vary substantially from galaxy to galaxy, and are in general steeper than the value of $\sim 1.4$ derived for large galaxies by \cite{ken98}. Further, as for the globally averaged quantities, the \ssfr\ at 400 pc resolution is in general lower than that predicted by the \cite{ken98} relation, with the deviation decreasing with increasing \sgas. Our computation of \ssfr\ uses a calibration that assumes solar metallicity and a standard Salpeter IMF, similarly, the \sgas\ we use is not corrected for the molecular gas density. Incorporating corrections for molecular gas and low metallicity will increase the deviation from the \cite{ken98} relation. Conversely, truncating the IMF at the high mass end would decrease the deviation from the \cite{ken98} relation. For the 5 galaxies for which a power law does not provide a good parametrization of the (\ssfr,\sgas) relation, there are substantial offsets between the UV bright regions and the HI high column density maps. Four of these five galaxies have HI masses near the lower end of our sample distribution, while the remaining galaxy has a large central HI hole. We have 200 pc resolution images for 10 of our galaxies. At this resolution, the offsets between the peaks in the HI and UV images are more pronounced, and a power law parametrization is possible for only 5 of the 10 galaxies.
\end{abstract}


\section{Introduction}
\label{sec:int}

   Understanding the physical processes that govern the conversion of interstellar gas into stars is crucial for a host of fundamental problems: galactic structure, the nature of the Hubble sequence, galaxy formation and evolution, and chemical evolution of the ISM. Since these processes, viz. how baryons accrete into dark matter halos, as well as subsequent star formation and feedback from star formation are complex and poorly understood, empirical relations between the gas column density (\sgas) and the star formation rate (\ssfr) surface density play an important role in semi-analytical models and simulations \citep[e.g.][]{spr05}. These empirical relations, or star formation ''recipes'' are generally derived from observations of large spiral galaxies. On the other hand, since in hierarchical models of galaxy formation, small objects form  first, and merge together later to form large galaxies, the star formation laws governing small, gas rich, galaxies are of particular interest. Gas in the smallest galaxies does not settle into a thin, dynamically cold disk, and the shallow potential wells of these galaxies would make their ISM more susceptible to disruption due to energy input from star formation. As such, it seems probable that star formation in the smallest galaxies proceeds in a manner that is different from that in large spirals. To the extent that the nearby extremely faint dwarf irregular galaxies could be regarded as proxies for galaxies in the early universe, it is interesting to look for correlations, if any, between the gas density and the star formation rate in them. In this paper we use HI data from the GMRT based FIGGS survey \citep{ay08} and publicly available \emph{GALEX} data to look for star formation recipes in a sample of extremely faint dwarf irregular galaxies. For ease of comparison with the existing large body of work on star formation recipes for galaxies, we examine in detail two particular recipes, (1)~the existence of a threshold column density below which star formation is quenched and (2)~a power law relation between the gas column density and the star formation rate density above this threshold.

   It is has long been suggested that there is a threshold column density below which star formation is quenched \citep{t64, spi68, qui72}. Cold gas in a thin rotating disk is unstable to gravitationally collapse above a critical column density \citep{safronov60,t64}. In large spiral galaxies therefore, such a threshold might be related to global disk instabilities. Sharp thresholds to  the H$\alpha$ emission have indeed been observed in normal spirals \citep{ken89,mar01}, leading credence to the idea that  a threshold for star formation does in fact exist. Our sample galaxies are unlikely to have thin, dynamically cold disks, since for dwarfs in this luminosity range the gas velocity dispersion is typically comparable in magnitude to the rotation velocity \citep{young03,begum03,begum04,begum06}. A threshold for star formation in dwarf galaxies may still exist as a consequence of a critical amount of dust shielding required for molecular gas to form \citep{ski87}.

      The suggestion that there is a power law relation between \ssfr~ and \sgas~ has a similarly long history. \cite{s59} related the volume densities of young stars and gas in the Galactic disk,viz. $\rm {\rho_{SFR} = a~\rho{_{gas}^n}}$ (the ``Schmidt law''). A more convenient parametrization for external galaxies is in terms of the surface densities, viz. $\rm {\Sigma_{SFR} = A~\Sigma{_{gas}^N}}$. There have been several measurements of the power law index {\sl N}, derived from observations of various samples and using various tracers of the star formation and the gas density. Kennicutt \citep{ken97} compiled a summary of the distribution of derived  values of {\sl N}, which had a broad peak between {\sl N}~=~0.8 and {\sl N}~=~2.5, and a full range of {\sl N}~=~$\pm$3.5. He concluded that the large dispersion could be only be partly attributed to observational factors, such as beam smearing effects or fitting to only the atomic or molecular gas densities. Much of the scatter is real, being caused by such factors as deviations from a power law or spatial variations in {\sl N} across a galaxy.  Kennicutt \citep{ken89, ken98} also investigated the possibility of a composite Schmidt law, using data ranging from normal spiral galaxies (star formation traced using H$\alpha$, gas density traced using both HI and CO observations) to circumnuclear starburst galaxies (star formation traced using Far Infrared, gas density traced using CO observations, HI gas being negligible), and arrived at the widely used \citep[for eg.~][~etc.]{spr00,nag04,kru07} ``Kennicutt--Schmidt'' law:
\begin{center}
\begin{equation}
\rm {\Sigma_{SFR}=(2.5\pm0.7)\times10^{-4}\Bigg(\frac{\Sigma_{gas}}{1M_\odot~pc^{-2}}\Bigg)^{1.4\pm0.15}M_\odot~yr^{-1}kpc^{-2}}
\label{eqn:law}
\end{equation}
\end{center}
\noindent We use this as a template to compare with the relations that we find in our sample of extremely faint dwarfs.

    As mentioned above, one point of departure from earlier studies is the fact that our sample is composed of extremely faint dwarfs. Another is the spatial resolution that we use. Most classical studies of star formation used either globally averaged surface densities or
azimuthally averaged radial profiles. This may be relevant in situations where, for e.g. global processes like large scale disk instabilities play a role in controlling star formation. For dwarf galaxies on the other hand, the low gas densities result in 
inefficient cloud formation as compared to inner part of spirals 
\citep{don03, li05}. Local processes are expected to dominate in dwarf galaxies  \citep{elm06}, and star formation is observed in HI clouds or complexes where the average gas density is much below the 
Toomre critical density \citep[eg.][]{deB06}. We hence study the small scale (i.e. $\sim$400 pc and $\sim$ 200 pc) relation between the gas and star formation rate in our sample galaxies by doing ``pixel by pixel'' correlations, in addition to looking at the correlation between globally averaged quantities. We also depart from many earlier studies in that we use only HI data to determine the gas column density, i.e. the contribution of the molecular gas is ignored. CO is notoriously difficult to detect in dwarf galaxies \citep[eg.][]{taylor98}, and the CO to H$_2$ conversion factor in dwarf galaxies may be substantially different from that in our own galaxy \citep[eg.][]{madden97,israel97}. Further the calibration we use to convert the \emph{GALEX} UV flux into a star formation rate assumes a standard Salpeter IMF with solar metallicity. We discuss the implications of these assumptions for our results in Sec.~\ref{sec:dis}.

\section{Sample and data analysis}
\label{sec:obs} 

Our  sample consists of 23 galaxies drawn from the FIGGS HI 21cm survey \citep{ay08}, for which there is publicly available \emph{GALEX} data.
The galaxies are listed in Table~\ref{tab:samp}; the columns in the table are: 
Column(1)~the galaxy name, 
Columns(2)\&(3)~the equatorial coordinates (J2000), 
Column (4)~the absolute blue magnitude (corrected for galactic extinction), 
Column(5)~the distance in Mpc, 
Column(6)~the group membership of the galaxy.
All of this data has been taken from \citet{ay08}.
Column(7)~the de Vaucouleurs (25 mag/arcsec$^2$) diameter of the optical disk.  For dwarf low surface brightness galaxies from the KK lists (KK14, KK65, KK144, KKH98), the diameters
correspond to the Holmberg system (~26.5 mag arcsec$^{-2}$).
Column(8)~the optical axis ratio. Data for column (7) and (8) have been taken from taken from \citet{kar04}.
Column(9)~the assumed inclination angle. The data is from \citet{ay08}, except for the starred galaxies, for which the inclination angle is measured from the coarsest resolution HI maps.

\begin{table*}
\begin{center}
\caption{ The sample}
\label{tab:samp}
\begin{tabular}{|lcccccccc|}
\hline
Galaxy&$\alpha$ (J2000)&$\delta$ (J2000)&M${\rm{_{B}}}$&Dist&Group&a&b/a&i$_{HI}$\\
      &(h~m~s)&($^\circ$~$^\prime$~$^{\prime\prime}$)&(mag)&(Mpc)&~&($^\prime$)&~&(deg)\\
\hline
\hline 
And IV&00 42 32.3&+40 34 19&$-$12.23&6.3&Field&1.3$~~$&0.77&55.0$~~$\\
UGC 685&01 07  22.3&+16 41 02&$-$14.31&4.5&Field&1.4$~~$&0.71&36.0$~~$\\
KK 14&01 44  42.7&+27 17 16&$-$12.13&7.2&N672&1.6$^+$&0.37&45.0$~~$\\
UGC 3755&07 13 51.8&+10 31 19&$-$14.90&6.96&Field&1.7$~~$&0.59&46.0$~~$\\
DDO 43&07 28 17.4&+40 46 11&$-$14.75&7.8&Field&1.3$~~$&0.69&30.0$~~$\\
KK 65&07 42 32.0&+16 33 40&$-$14.29 &7.62&Field&0.9$^+$&0.56&47.0$~~$\\
UGC 4459&08 34 06.5&+66 10 45&$-$13.37&3.56&M81&1.6$~~$&0.87&30.0$~~$\\
UGC 5186&09 42 59.1&+33 16 00&$-$12.98&6.9&Field&1.3$~~$&0.23&53.3$^*$\\
UGC 5209&09 45 04.2&+32 14 18&$-$13.15&6.7&Field&0.9$~~$&0.96&25.4$^*$\\
UGC 6456&11 28 00.6&+78 59 29&$-$14.03&4.3&M81&1.5$~~$&0.53&65.0$~~$\\
UGC 6541&11 33 28.9&+49 14 14&$-$13.71&3.9&CVn~I&1.4$~~$&0.57&44.9$^*$\\
NGC 3741&11 36 06.4&+45 17 07&$-$13.13&3.0&CVn~I&2.0$~~$&0.55&68.0$~~$\\
DDO 99&11 50 53.0&+38 52 50&$-$13.52&2.6&CVn~I&4.1$~~$&0.37&50.1$^*$\\
E321-014&12 13 49.6&$-$38 13 15&$-$12.70&3.2&Cen A&1.4$~~$&0.43&56.5$^*$\\
KK 144&12 25 27.9&+28 28 57&$-$12.59&6.3&CVn~I&1.5$^+$&0.33&57.0$~~$\\
DDO 125&12 27 41.8&+43 29 38&$-$14.16&2.5&CVn~I&4.3$~~$&0.56&44.9$^*$\\
UGC 7605&12 28 38.9&+35 43 03&$-$13.53&4.43&CVn~I&1.1$~~$&0.73&40.0$~~$\\
UGC 8215&13 08 03.6&+46 49 41&$-$12.26&4.5&CVn~I&1.0$~~$&0.70&45.0$~~$\\
DDO 167&13 13 22.8&+46 19 11&$-$12.70&4.2&CVn~I&1.1$~~$&0.55&45.8$^*$\\
DDO 181&13 39 53.8&+40 44 21&$-$13.03&3.1&CVn~I&2.3$~~$&0.57&53.0$~~$\\
DDO 183&13 50 50.6&+38 01 09&$-$13.17&3.24&CVn~I&2.2$~~$&0.32&67.0$~~$\\
UGC 8833&13 54 48.7&+35 50 15&$-$12.42&3.2&CVn~I&0.9$~~$&0.89&26.0$~~$\\
KKH 98&23 45 34.0&+38 43 04&$-$10.78&2.5&Field&1.1$^+$&0.55&46.0$~~$\\
\hline
\hline
\end{tabular}
\end{center}
\begin{flushleft}
$^*$: value recomputed by us\\
$^+$: diameters correspond to the Holmberg system (~26.5 mag arcsec$^{-2}$)
\end{flushleft}
\end{table*}

\subsection{HI data}
\label{ssec:hidata}

The analysis of the HI data is described in detail in \citet{ay08}. 
For computing the spatially resolved star formation law, it is important that all galaxies be imaged at the same linear resolution. We hence re imaged the calibrated FIGGS HI visibility data to make CLEANed data cubes at a linear resolution as near 400 pc as possible using the task IMAGR in classic AIPS. Maps of the total integrated flux (``Moment 0'' maps) were made from the data cubes using the task MOMNT. The cut offs used in the MOMNT task was $\sim 2\sigma$, where $\sigma$ is the rms level in a single channel. The images were re-gridded to have a pixel size identical to that of the \emph{GALEX} UV images using the task HGEOM. 
For pixel to pixel comparison, the flux values were averaged over $n \times n$ pixel sized boxes, where $n$ is an integer chosen (independently for each galaxy), such that four such boxes approximately cover a circular area of diameter $\sim$400~pc.Each such box is the ''pixel'' we refer to when we say pixel by pixel analysis. For each such ''pixel'' the flux was  converted to column density values for HI along the line of sight, in units of M$_\odot$ pc$^{-2}$, assuming that the emission is optically thin; the face on column density was computed after correcting for the inclination angle using the values given in Table~\ref{tab:samp}. 

\begin{table}
\begin{center}
\caption{ The Parameters derived during analysis, for 400 pc resolution}
\label{tab:beam}
\begin{tabular}{|lcccc|}
\hline
Galaxy&synthesised beam&synthesised beam&Noise&a$_{sf}$\\
      &(arcsec$^2$)&(pc$^2$)&(mJy)&($^\prime$)\\
\hline
\hline
And IV&13.86$\times$12.71&423$\times$388&1.4&1.7\\
UGC 685&16.81$\times$15.98&367$\times$349&1.8&1.4\\
KK 14&13.38$\times$09.99&467$\times$349&2.1&1.0\\
UGC 3755&11.81$\times$11.07&398$\times$374&1.9&1.7\\
DDO 43&11.67$\times$9.53&441$\times$360&1.5&2.0\\
KK 65&11.49$\times$10.26&424$\times$379&1.3&1.0\\
UGC 4459&24.86$\times$21.09&429$\times$364&1.6&2.0\\
UGC 5186&12.14$\times$10.48&406$\times$351&1.1&1.1\\
UGC 5209&12.80$\times$10.70&416$\times$348&2.0&0.9\\
UGC 6456&19.38$\times$17.01&404$\times$355&2.9&2.0\\
UGC 6541&22.68$\times$21.36&429$\times$404&3.4&1.9\\
NGC 3741&28.21$\times$27.02&410$\times$393&2.4&2.2\\
DDO 99&33.26$\times$29.41&419$\times$371&2.9&3.6\\
E321-014&29.88$\times$21.68&464$\times$336&2.5&1.8\\
KK 144&15.99$\times$10.28&488$\times$314&2.0&1.6\\
DDO 125&34.78$\times$30.04&422$\times$364&3.4&3.3\\
UGC 7605&22.28$\times$14.84&479$\times$319&1.7&1.6\\
UGC 8215&19.02$\times$17.94&415$\times$391&2.4&1.0\\
DDO 167&20.66$\times$18.83&421$\times$383&3.4&1.7\\
DDO 181&29.77$\times$25.16&447$\times$378&2.7&2.7\\
DDO 183&27.42$\times$24.46&431$\times$384&2.2&2.5\\
UGC 8833&26.44$\times$25.17&410$\times$390&2.6&1.3\\
KKH 98&34.46$\times$31.63&418$\times$383&2.4&1.4\\
\hline
\hline
\end{tabular}
\end{center}
\end{table}

\subsection{UV data}
\label{ssec:uvdata}

 FUV and NUV images for our sample galaxies are publicly available on the \emph{GALEX} site. We present analysis based on the FUV band (1350-1750 $\rm{\AA}$) only, since a portion of the NUV band  (1750-2800 $\rm{\AA}$) lies outside the range where  the calibration for conversion between the UV flux and the star formation rate, i.e. eqn no.(\ref{eqn:cal}) below is valid. The results presented in
Section~\ref{sec:res} below are however not much affected by whether one uses the FUV or NUV data.  
First the geometry (orientation, pixel size and number of pixels) of the FUV maps were made identical to that of the HI maps using the task HGEOM in AIPS,  the images were then smoothed to the same resolution as the HI images (i.e. linear resolution of $\sim$400 pc). The HI contours overlayed on the FUV grey scales are shown in Figure~\ref{fig:olay}.
Then the FUV maps, which are background subtracted FUV images (with units of counts per second) were  converted into luminosity units using the calibration information provided at the \emph{GALEX} site as given below, and then corrected for galactic  extinction (using \citet{sch98}'s galactic extinction values in the direction of the particular galaxy, and using formulae given by \citet{car89} to extrapolate to the FUV band). No correction for internal extinction was made, since our sample galaxies are expected to be extremely dust poor. The validity and implications of this particular assumption are discussed in detail in Sec~\ref{sec:dis}.

\begin{center}
\begin{equation}
\rm {m_{GALEX}=-2.5\log(cps)}
\label{eqn:gal1}
\end{equation}

\begin{equation}
\rm {m_{AB}=m_{GALEX}+18.82}
\label{eqn:gal2}
\end{equation}
\end{center}

\noindent
The luminosity values thus obtained were converted to surface density of star formation, in units of M$_\odot$ yr$^{-1}$ kpc$^{-2}$ using the calibration given in \citet{ken98a} :

\begin{equation}
\rm {SFR(M_{\odot}~year^{-1})~=~1.4\times10^{-28}L_{\nu}(ergs~s^{-1}~Hz^{-1})}
\label{eqn:cal}
\end{equation}

\noindent
which is valid for the wavelength range viz. 1250-2500 $\rm{\AA}$ \citep[see,][]{ken98a}. The assumptions that go into deriving the above calibration are that the stars have solar metallicity and Salpeter IMF, and that the galaxy has had continuous star formation over time scales of 10$^8$ years or longer. The calibration is also sensitive to extinction by dust. From the well known luminosity-metallicity relationship \citep[for eg.][]{pilyugin01, lee06}, one would expect that the typical metallicity in our sample galaxies is substantially less than solar. We do not know the detailed star formation history for many of our sample galaxies, nor do we have information on what their IMF may be. We discuss these issues in some more detail in Section~\ref{sec:dis}, but this uncertainty in the application of this calibration to our sample galaxies should be borne in mind.

  As discussed above, a uniform optical diameter is not available for all the galaxies in our sample. For those galaxies for which the Holmberg diameter is available, the average UV flux at this diameter corresponds to a star formation rate of $\sim$ 1.85$\times$10$^{-4}$ M$_\odot$ yr$^{-1}$ kpc$^{-2}$. The globally averaged star formation
rate for all galaxies was hence computed to be the average star
formation rate inside the region where the SFR falls to 1.85$\times$10$^{-4}$ M$_\odot$ yr$^{-1}$ kpc$^{-2}$ (where the average SFR was calculated along ellipses in the smoothed FUV image having ellipticity b/a as given in Table~\ref{tab:samp}). Both \sgas~ and \ssfr~ were averaged over identical regions for each galaxy.

Table~\ref{tab:beam} summarizes the parameters derived during the analysis. The columns are as follows: 
(1)~the galaxy name, 
(2)~the synthesized beam sizes of the HI maps (which are also the smoothed resolution of the  FUV images) in arcseconds, 
(3)~the synthesised beam sizes in parsecs, 
(4)~the rms noise per channel of the cleaned HI data cubes,
(5)~the major axis in arcseconds of the ellipse, defining the ''star forming region'', as described above.

\begin{figure*}
\psfig{file=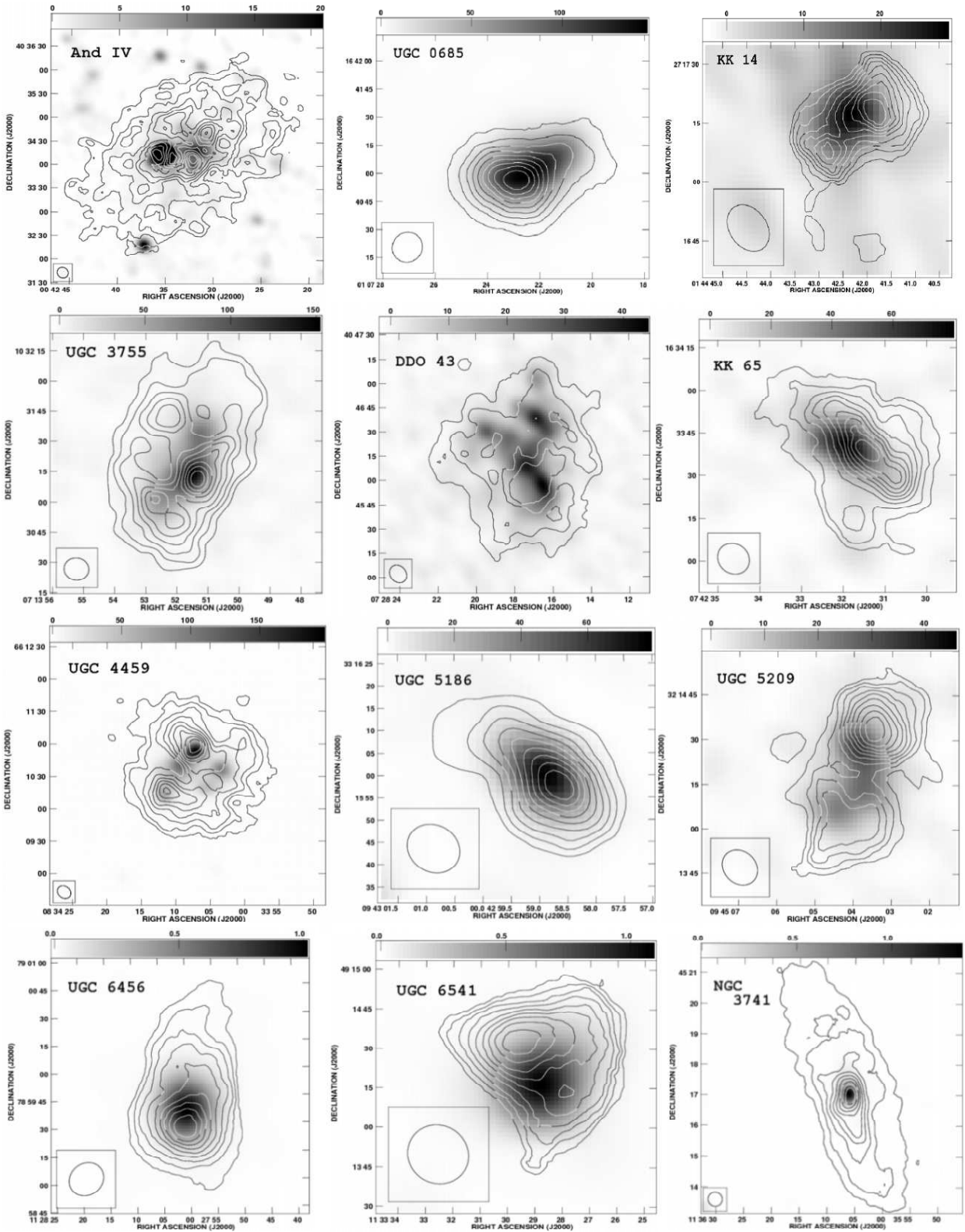,width=6.5truein}
\caption{HI contours overlayed on gray scale images of smoothed FUV intensity maps for each individual galaxy. The contours are equally spaced at intervals of 10 percent of the peak value.}
\label{fig:olay}
\end{figure*}

\begin{figure*}
\addtocounter{figure}{-1}
\psfig{file=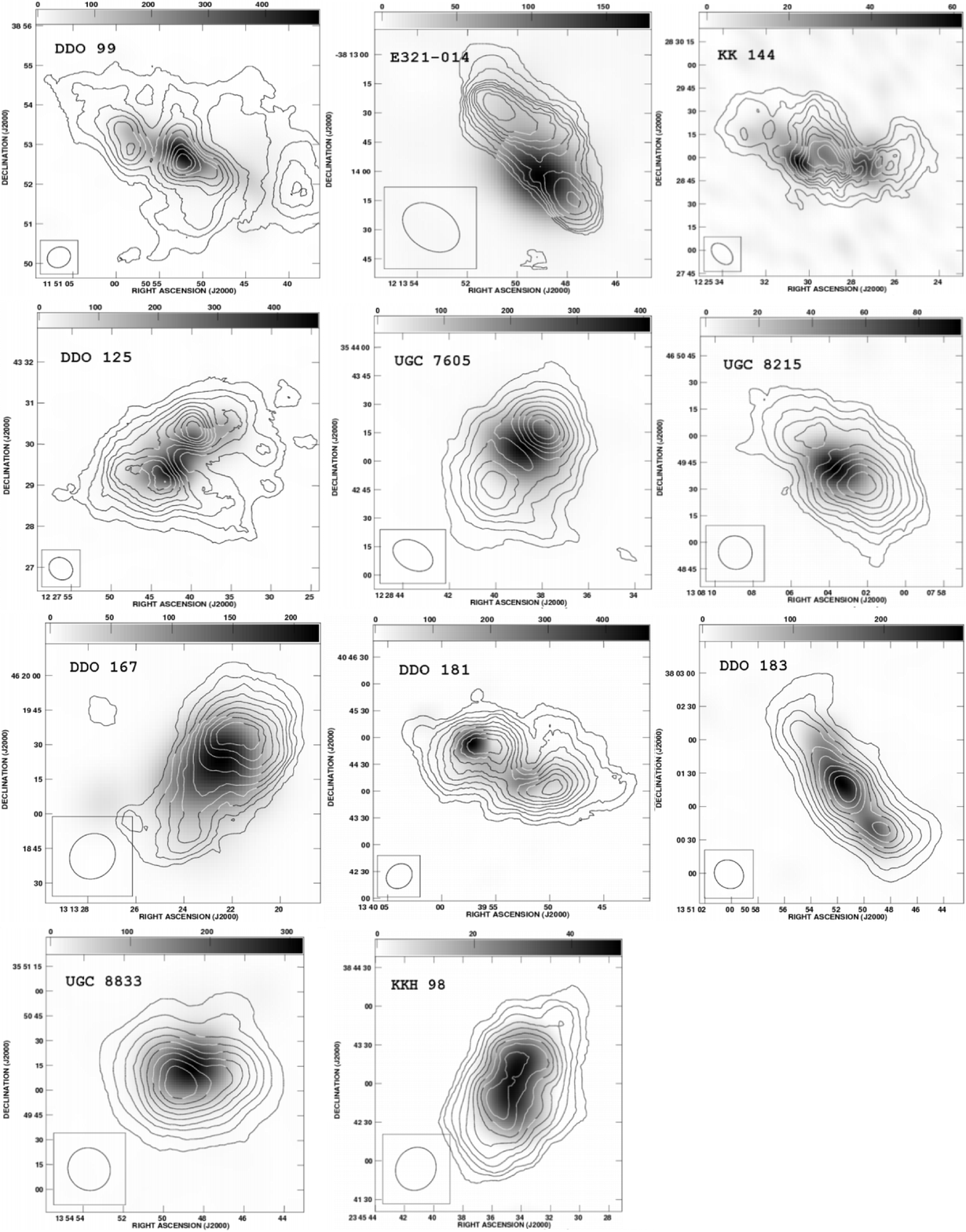,width=6.5truein}
\caption[Continued]{({\it{continued}})}
\label{fig:olaya}
\end{figure*}

\begin{figure*}
\psfig{file=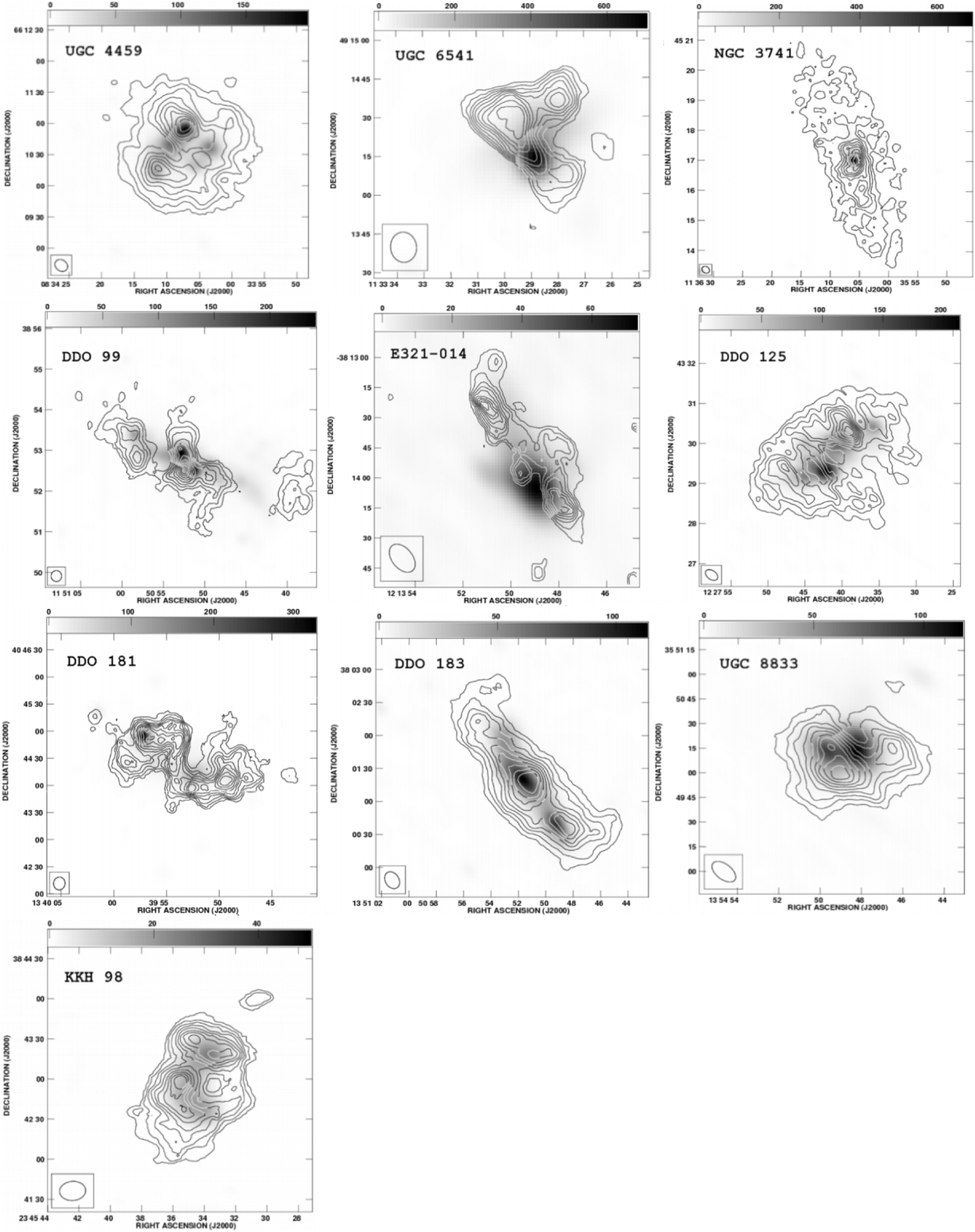,width=6.5truein}
\caption{HI contours overlayed on gray scale images of smoothed FUV intensity maps for each individual galaxy (linear resolutions of $\sim$200 pc). The contours are equally spaced at intervals of 10 percent of the peak value.}
\label{fig:olay1}
\end{figure*}

For 10 galaxies in the sample, it was possible to make maps at a linear resolution of $\sim$200 pc, and so a pixel by pixel comparison of FUV and HI data was carried out for these galaxies at this higher resolution. All other processing was done in a manner analogous to the procedure discussed above. The relevant details are listed in ~\ref{tab:beam1}, whose columns are as follows: 
(1)~the galaxy name, 
(2)~the synthesized beam sizes for the HI maps (which are also
the smoothed resolution of the FUV images) in arcseconds, 
(3)~the synthesised beam sizes in parsecs, 
(4)~the rms noise per channel of the cleaned HI data cubes.
The HI contours overlayed on the FUV grey scales are shown in
Figure~\ref{fig:olay1}.

\begin{table}
\begin{center}
\caption{ The Parameters derived during analysis, for 200 pc resolution}
\label{tab:beam1}
\begin{tabular}{|lccc|}
\hline
Galaxy&synthesised beam&synthesised beam&Noise\\
      &(arcsec$^2$)&(pc$^2$)&(mJy)\\
\hline
\hline
UGC 4459&12.76$\times$10.47&220$\times$181&1.4\\
UGC 6541&11.71$\times$9.50&221$\times$180&2.0 \\
NGC 3741&14.94$\times$12.32&217$\times$179&1.8\\
DDO 99&16.37$\times$14.84&206$\times$187&2.6  \\
E321-014&16.03$\times$10.09&249$\times$157&2.0\\
DDO 125&20.64$\times$13.19&250$\times$160&2.7 \\
DDO 181&14.33$\times$12.22&215$\times$184&2.2 \\
DDO 183&14.17$\times$10.90&223$\times$171&1.7 \\
UGC 8833&16.44$\times$9.21&255$\times$143&1.6 \\
KKH 98&18.76$\times$14.39&227$\times$174&2.4  \\
\hline
\hline
\end{tabular}
\end{center}
\end{table}

\section{Results}
\label{sec:res}

\subsection{Global Averaged Quantities}
\label{ssec:global}

\begin{figure}
\psfig{file=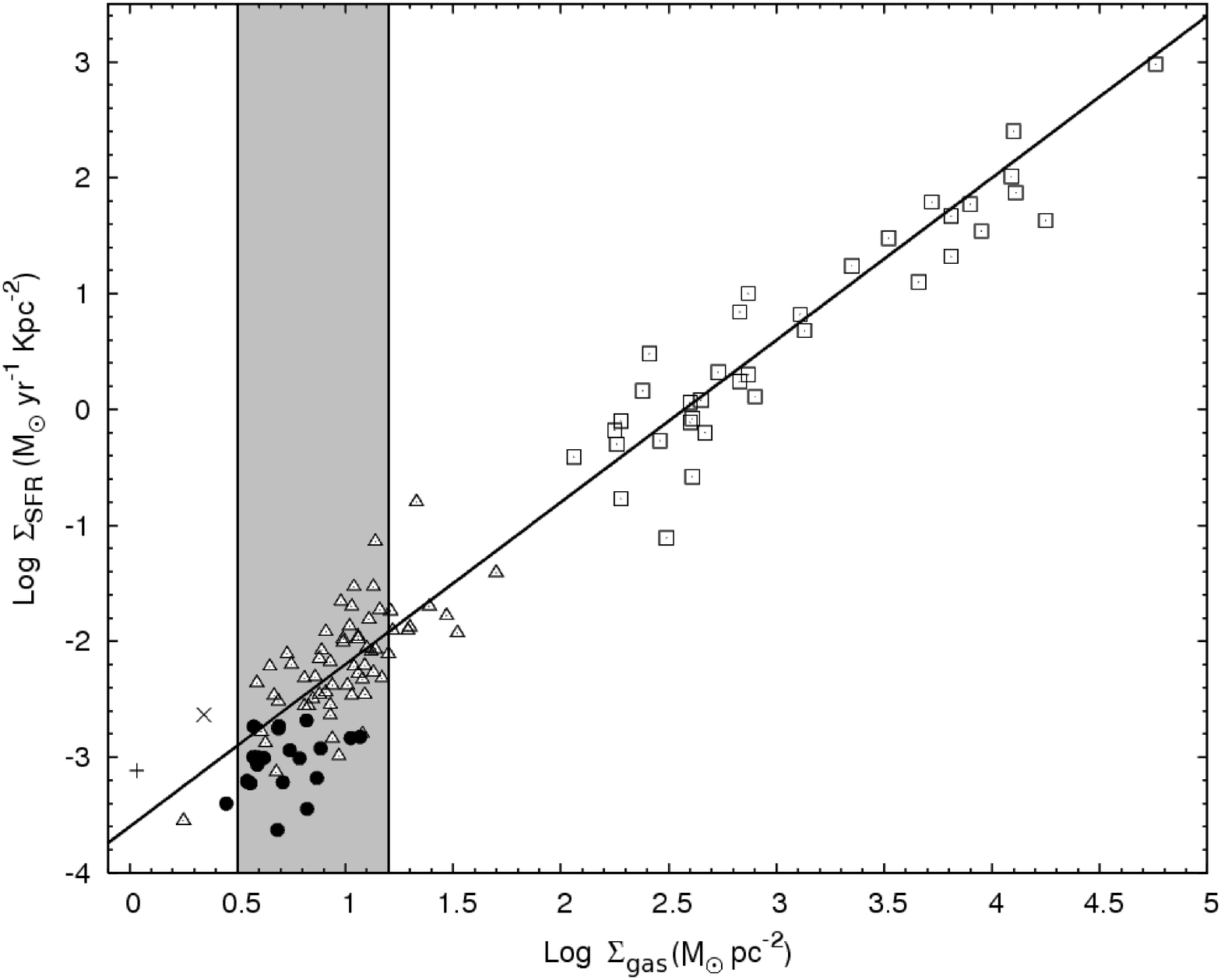,width=3.5truein}
\caption{Scatter plot of the  disk-averaged SFR per unit area and gas surface density. The filled symbols represent data from the present sample. The open symbols are from \citet{ken98}, with triangles for data from normal spiral galaxies,and open squares for data from circumnuclear starburst galaxies. The solid line is the best fit power law relation from \citet{ken98} ~(i.e. eqn.~\ref{eqn:law} above). The cross sign represents the data from UGC 6541, the plus sign represents the data from E321-014. These are the only two galaxies in our sample which lie above the \citet{ken98} relation inspite of having gas surface densities much less than the expected ''threshold density'' values. The shaded region covers various estimates of the ``threshold density'' tabulated in \citet{ken89}.}
\label{fig:tot}
\end{figure}

Figure~\ref{fig:tot} shows the relationship between the disk-averaged SFR and HI gas surface density for the galaxies in our sample, along with those from  \citet{ken98}. The scatter in the disk-averaged data from the galaxies in our sample is too large, and the range of star formation rates and gas densities covered is too small, to provide any meaningful constraints on the power law slope. The fit shown in Figure~\ref{fig:tot} is instead taken from \cite{ken98}. The shaded area covers various estimates of the ``threshold density'' as tabulated in \cite{ken89,mar01}. As can be seen our sample galaxies (1)~have gas densities that are generally around or below the expected ''threshold density'' and (2)~have star formation rates below that predicted by the \cite{ken98} relation. These two facts are probably not independent. 
Two galaxies, viz. E321-014 and UGC 6541 lie significantly above the \cite{ken98} relation inspite of being sufficiently below the threshold gas density region, and we discuss them in more detail below.

\begin{figure}
\psfig{file=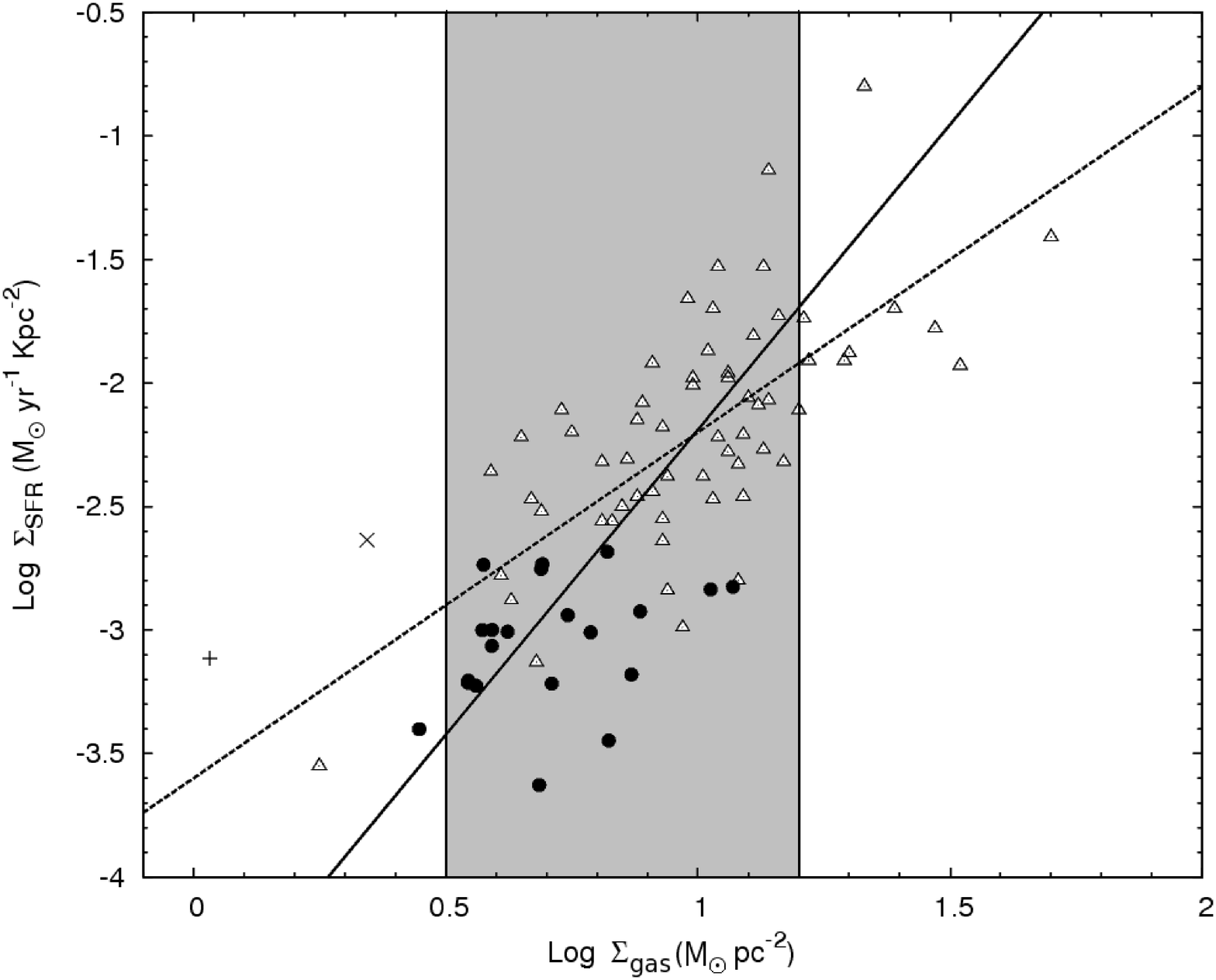,width=3.5truein}
\caption{Scatter plot of the  disk-averaged SFR per unit area and gas surface density, for dwarf and spiral galaxies only. The solid line is the best fit power law relation for spiral galaxies (from \citet{ken98}, with a slope of 2.47). The dashed line represents the ``Kennicutt-Schmidt'' law, i.e. eqn.~\ref{eqn:law} above. All other symbols are identical to those in Figure~\ref{fig:tot}.}
\label{fig:spc}
\end{figure}

The solid line in Fig.~\ref{fig:spc} shows the \citet{ken98} fit to the data for spiral galaxies alone. As can be seen, the dwarf galaxy data is in reasonable agreement with this fit. As such, it appears that dwarf and spiral galaxies have a steeper dependence of SFR on the gas density than predicted by the Kennicutt-Schmidt law.

\subsection{Small scale relations}
\label{ssec:400pc}

\begin{figure*}
\psfig{file=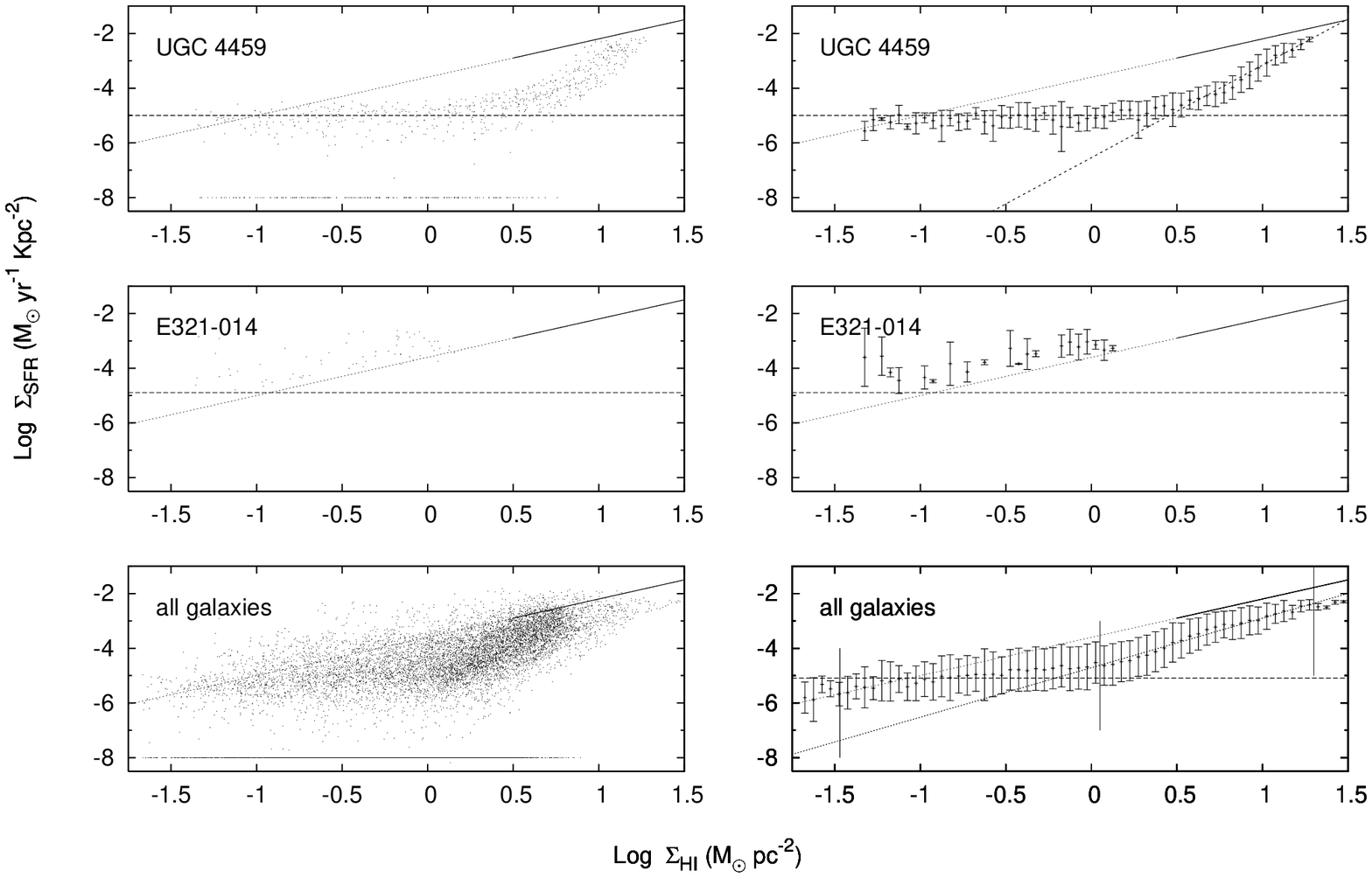,width=6.5truein}
\caption{The pixel by pixel comparison of gas density and SFR at 400 pc resolution. The first two rows are the plots for representative galaxies UGC 4459 (non-deviant) and E321-014 (deviant) taken from Figure~\ref{fig:plots}. The last row is the combined plot for data points taken from all galaxies. The vertical lines in this plot mark the ranges over which the flat (for low \sgas\ ) portion and the power-law (for high \sgas\ ) portion have been fitted. See Section~\ref{ssec:400pc} and Appendix~\ref{sec:ap}.}
\label{fig:tp}
\end{figure*}

    The ``pixel by pixel'' (see Section~\ref{ssec:hidata} for the definition of a pixel) comparison of \sgas\  and \ssfr\ for individual galaxies at 400~pc resolution are shown in Figure~\ref{fig:plots}. Plots for the representative galaxies UGC~4459 and E321-014, along
with the combined scatter plot for all the galaxies in the sample are shown in Figure~\ref{fig:tp}. The left panel for each galaxy shows the scatter plot, each point denoting a pixel. The right panel shows the binned data (0.05 dex sized bins were used along the x-axis), with the scatter (1$\sigma$) in each bin denoted by errorbars. The dotted line is the ''Kennicutt--Schmidt'' law (\citet{ken98}; eqn.~\ref{eqn:law} above), with the bold solid portion of the line lying above the lowest estimated of the threshold density listed in \citep{ken89}. For most galaxies, the scatter plot has a flat portion at low \sgas\  where the \ssfr\  density is constant, followed by an approximately power-law increase in \ssfr\ with \sgas. 
The horizontal dotted line in each panel corresponds to be sensitivity level of the \emph{GALEX} images; as can be seen the flat portion of the scatter plot basically corresponds to the region where \ssfr\  falls below the level that can be measured from the current data. A straight line was fitted to the binned data iteratively, considering only those points which lie above the FUV sensitivity limit and above the surface gas density value where the straight line meets the FUV sensitivity limit line, starting with an initial guess. For some ''deviant'' galaxies (viz. E321-014, UGC 6541, UGC 5209, KK 14 and DDO 43) this procedure did not  converge. For these galaxies the HI and FUV peaks can be seen in Figure~\ref{fig:olay} to be offset, making it natural that the scatter plot shows deviations from a simple power law relation. It is worth noting that the galaxy with the smallest HI mass (see Table~\ref{tab:res}), viz. E321-014 is the most deviant, with the FUV emission being considerably offset from the HI emission. This galaxy (along with UGC~6541) were also outliers in the global relation (Fig.~\ref{fig:tot}). Another interesting feature seen in Figure~\ref{fig:plots} is that while the observed \ssfr\ generally lies below that predicted by the fiducial \cite{ken98} relation, at high gas column densities, the observed \ssfr\ begins to approach the predicted rate. The fact that \ssfr\  is generally lower than that predicted by the \cite{ken98} relation was also seen earlier for the global averaged values (Figure~\ref{fig:tot}).

Quantitative results from the pixel by pixel correlation are listed in Tables~\ref{tab:res}. The columns are:
(1)~Galaxy name, 
(2)~the  derived HI mass taken from \citet{ay08}, 
(3)~sensitivity limit of FUV data (i.e. the RMS noise in the background emission),
(4)~the best fit power law index,
(5)~coefficient of the power law fit,
(6)~the range of bins along the gas surface density axis over which the straight line representing the power law part was fitted.

\begin{table*}
\begin{center}
\caption{The derived quantities, for 400 pc resolution}
\label{tab:res}
\begin{tabular}{|lccccc|}
\hline
Galaxy&Total M$_{HI}$&sensitivity limit of FUV&power law index&coefficient in log&range of fit\\
\\
      &[10$^6$ M$_\odot$]&[log(M$_\odot$ yr$^{-1}$ Kpc$^{-2}$)]&~\\
\hline
\hline
And IV&205.19&$-$4.89&2.3$\pm$0.1&$-$5.4$\pm$0.1&[0.225,0.925]\\
UGC 685&56.15&$-$4.91&1.45$\pm$0.06&$-$4.5$\pm$0.8&[-0.275,1.475]\\
KK 14&21.93&$-$4.04&~\\
UGC 3755&41.30&$-$3.87&2.6$\pm$0.1&$-$4.9$\pm$0.1&[0.425,1.125]\\
DDO 43&203.02&$-$4.12&~\\
KK 65&34.38&$-$4.14&1.7$\pm$0.1&$-$4.4$\pm$0.1&[0.175,1.025]\\
UGC 4459&64.2&$-$5.0&3.38$\pm$0.08&$-$6.54$\pm$0.09&[0.475,1.275]\\
UGC 5186&15.66&$-$4.53&1.62$\pm$0.07&$-$3.98$\pm$0.03&[-0.275,0.825]\\
UGC 5209&21.10&$-$4.36&~\\
UGC 6456&43.89&$-$4.60&1.8$\pm$0.1&$-$3.70$\pm$0.06&[-0.475,0.925]\\
UGC 6541&9.65&$-$4.61&~\\
NGC 3741&130.0&$-$5.35&4.4$\pm$0.2&$-$6.3$\pm$0.2&[0.225,0.875]\\
DDO 99&52.42&$-$5.15&2.6$\pm$0.2&$-$4.8$\pm$0.1&[-0.125,0.875]\\
E321-014&3.13&$-$4.90&~\\
KK 144&81.15&$-$4.57&1.46$\pm$0.06&$-$4.50$\pm$0.06&[-0.025,1.025]\\
DDO 125&31.87&$-$5.37&2.41$\pm$0.07&$-$4.66$\pm$0.04&[-0.275,0.775]\\
UGC 7605&22.29&$-$5.13&1.60$\pm$0.06&$-$3.69$\pm$0.05&[-0.875,0.975]\\
UGC 8215&21.41&$-$5.26&3.1$\pm$0.2&$-$5.7$\pm$0.1&[0.175,0.825]\\
DDO 167&14.51&$-$5.19&1.8$\pm$0.2&$-$3.8$\pm$0.1&[-0.675,0.875]\\
DDO 181&27.55&$-$5.38&2.6$\pm$0.1&$-$4.83$\pm$0.05&[-0.175,0.725]\\
DDO 183&25.90&$-$5.58&1.88$\pm$0.04&$-$4.26$\pm$0.03&[-0.675,0.725]\\
UGC 8833&15.16&$-$5.06&2.25$\pm$0.09&$-$5.08$\pm$0.07&[-0.075,1.075]\\
KKH 98&6.46&$-$4.79&3.1$\pm$0.1&$-$4.74$\pm$0.04&[0.025,0.475]\\
\hline
\hline
\end{tabular}
\end{center}
\end{table*}

\begin{figure*}
\psfig{file=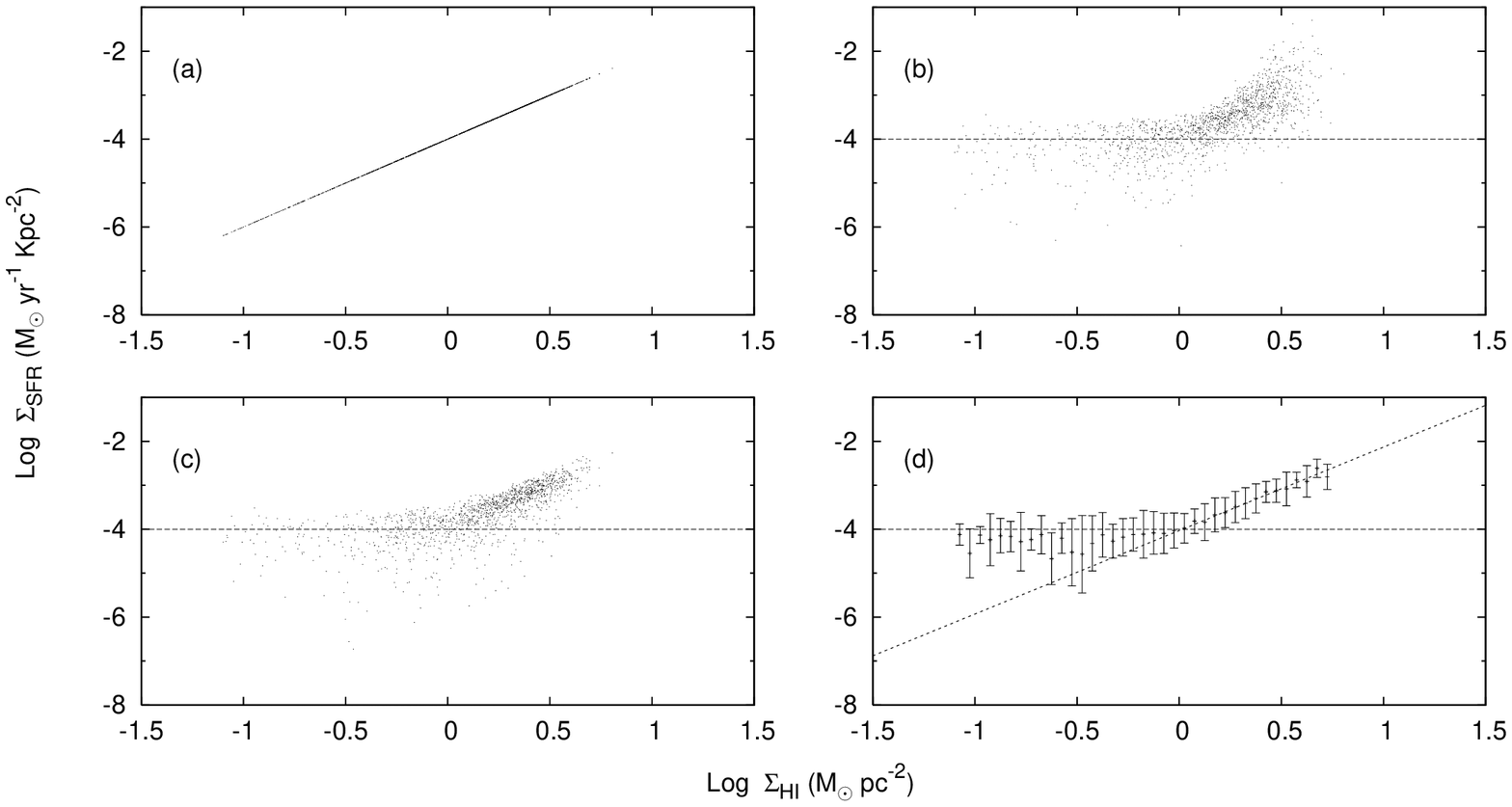,width=6.5truein}
\caption{Simulated ``pixel by pixel'' correlation for a power law relation between the star formation rate density and the gas density. Panel~[a] shows the correlation in the absence of noise, while in panel~[b] noise has been added. The noise level (1$\sigma$) is indicated by the horizontal line. As can be seen the scatter plot is flat at SFR levels below the sensitivity threshold.In addition, the power law coefficient is allowed to vary by 50\% from pixel to pixel. Unlike what is observed for the real data (Fig.~\ref{fig:plots},\ref{fig:plots200}) the scatter increases towards high column densities. In panel~(c) noise is added to the data as before, but instead of varying the power law index the coefficient of the power law is however allowed to vary by 50\%. This results in a better match to the real data. Panel~(d) shows the same data as in panel~(c) after binning. The fit line recovers the input values to the simulation. See the text for more details.}
\label{fig:sim}
\end{figure*}

As can be seen from the table, the parameters of the power law fit vary substantially from galaxy to galaxy. Nonetheless, for all galaxies the star formation appears to continue smoothly until one reaches the sensitivity limit of the observations, i.e. there does not seem to be any evidence for a ``threshold density'' below which the star formation is completely quenched. Figure~\ref{fig:sim} illustrates this point. Panel (a)~shows simulated data in which 
\ssfr\ is related to the \sgas\ by a power law with index $2$
and coefficient $-4$. These values have been chosen to be similar to the observed power law indices and coefficients. In panel~(b) noise is added, as seen in the real data, this leads to a flattening of the scatter plot for \ssfr\  that are below the sensitivity threshold. The horizontal line shows the $1\sigma$ noise level. In addition, the power law coefficient is allowed to vary by 50\% from pixel to pixel. As can be seen the scatter in the plot increases towards high column densities, unlike that seen in the real data. In panel~(c) noise is added to the data as before, but the power law index is held constant at the value of 2. The coefficient of the power law is however allowed to vary by 50\%. This results in a better match to the real data than varying the index of the power law. Panel~(d) shows the same data as in panel~(c) after binning; the bins are of the same size as was used for actual data. A linear relation to the fit data was done using the same procedure as was followed for the real data, the fit parameters were a coefficient of $-4.03 \pm 0.7$ and a power law index of $1.9 \pm 0.1$, i.e. the input values to the simulation were recovered.

The combined data for all galaxies (at 400 pc resolution),is shown in Figure~\ref{fig:tp}. As for the individual galaxies, the plot shows
a relatively flat portion at low \shi\ and a power law like behavior
at high \shi. The flat portion has log(\ssfr)$ = -5.09$ (for \shi\  varying between -1.47 and 0.05), and the slope at the high \shi\ end (\shi\ varying between 0.05 and 1.31) is 1.81. The vertical arrows
in Fig.~\ref{fig:tp} show the range over which the fits have been
made. A few points at both extremes have been omitted from the fit,
because each set arises from a single galaxy (viz. NGC~3741 at the
low \shi\ end and UGC~685 at the high \shi\ end). Since the UV images of the different galaxies have different sensitivity levels, the sloping part of the plot contains data from both the flat and sloping parts of the individual galaxy scatter plots. As such, the average slope may not physically meaningful. However, if the galaxies for which there are offsets between the HI and the UV peaks, and the flat portions of the individual scatter plots are excluded from the fit, the average slope does not change significantly. As such the above fit can be treated as representative (in average sense) of  the entire sample.

The linear resolution, viz. 400 pc that we have been working at is still large compared to the scales on which one might expect star formation to occur and the effects of feedback from star formation to be dominant. As mentioned in Section~\ref{sec:obs}, for 10 galaxies in the sample, it was possible to make maps at a linear resolution of $\sim$200 pc (Figure~\ref{fig:olay1}). The scatter plot of the SFR  vs. the gas column density for these galaxies is shown in  Figure~\ref{fig:plots200}. Where possible, a power law fit was made to the binned data in the same manner as for the 400 pc resolution data. The parameters from this fit are given in Table~\ref{tab:res200}, the columns are the same as those in Table~\ref{tab:res}, except that the values correspond to those measured at a 200 pc resolution. For 3 of  the ten galaxies (viz. DDO99, DDO181 and KKH98) a power law relation could be found at a 400 pc resolution, but not at 200 pc. In two of these cases (viz DDO99 and DDO181) this is because offsets between the HI and UV peaks which were not prominent at the 400 pc resolution become so at the 200 pc resolution. For the third galaxy (KKH98) at high resolution there are too few points to make a good fit. For two more galaxies (viz. UGC6451 and E321-04) no power law fit could be made either at 400 pc or 200 pc resolution and a comparison of Figures~(\ref{fig:olay}) and ~(\ref{fig:olay1}) shows that the offsets between the HI and UV seen in the 400 pc resolution images have become more pronounced at 200 pc resolution. For the remaining 5 galaxies meaningful power law fits are possible for both resolutions, and for 4 of these 5 (i.e. except DDO183) the fit slope is shallower and the power law coefficient is larger at 200 pc resolution than at 400 pc. In the case of DDO183, a small offset between the UV and HI images can be seen at 200 pc resolution while the two are quite well aligned at the 400 pc resolution. Similarly for UGC8833, even though a power law fit could be obtained for the 200 pc resolution data, a comparison of the overlay images show that what at 400 pc resolution appeared to be coincident HI and FUV  emission, actually appears offset at 200 pc resolution. For all the galaxies for which a meaningful power law fit can be made, the observed SFR is lower than that predicted by the \cite{ken98} relation, although again the difference in general diminishes with increasing gas column density.

\begin{table*}
\begin{center}
\caption{The derived quantities, for 200 pc resolution}
\label{tab:res200}
\begin{tabular}{|lccccc|}
\hline
Galaxy&Total M$_{HI}$&sensitivity limit of FUV&power law index&coefficient in log&range of fit\\
\\
      &[10$^6$ M$_\odot$]&[log(M$_\odot$ yr$^{-1}$ Kpc$^{-2}$)]&~\\
\hline
\hline
UGC 4459&64.2&$-$4.75&2.75$\pm$0.09&$-$5.98$\pm$0.09&[0.475,1.375]\\
UGC 6541&9.65&$-$4.27&~\\
NGC 3741&130.0&$-$5.20&3.9$\pm$0.1&$-$6.1$\pm$0.1&[0.225,1.025]\\
DDO 99&52.42&$-$4.93&~\\
E321-014&3.13&$-$4.64&~\\
DDO 125&31.87&$-$5.17&1.50$\pm$0.04&$-$4.26$\pm$0.02&[-0.575,0.925]\\
DDO 181&27.55&$-$5.18&~\\
DDO 183&25.90&$-$5.31&2.6$\pm$0.1&$-$5.26$\pm$0.09&[0.025,0.975]\\
UGC 8833&15.16&$-$4.81&1.37$\pm$0.03&$-$4.38$\pm$0.03&[-0.275,1.225]\\
KKH 98&6.46&$-$4.59&~\\
\hline
\hline
\end{tabular}
\end{center}
\end{table*}

\section{Discussion}
\label{sec:dis}

      To summarize our results, we find in the preceding section that (1)~the globally averaged gas density in our sample galaxies lies below most estimates of the star formation rate, and the observed \ssfr\  is also lower than that estimated from the \cite{ken98} relation. (2)~The data is better fit by the steeper slope found for the spirals only sample by \cite{ken98}. (3)~For most (18/23) galaxies \ssfr\  can be parametrized to have a power law dependence on \sgas\  when both parameters are measured on 400 pc scales, however the coefficient and index of the power law varies substantially from galaxy to galaxy,(4)~the index of the power law is in general steeper than the value of 1.4 for the \cite{ken98} relation. The observed SFR rate is in  general lower than that predicted by this relation, with the discrepancy decreasing at the highest gas column densities. (5)~At 400 pc resolution the SFR continues to fall smoothly until one reaches the sensitivity limit of our observations, and there is no evidence for a ``threshold density'' below which star formation is completely cut off, (6)~at 200 pc resolution, offsets between the sites of current star formation and the locations where the gas density peaks become more pronounced, and the SFR can be parametrized as having a power law dependence on the gas density for only 5/10 galaxies, and (7) for the majority of these galaxies (4/5) the power law index measured at 200 pc is flatter than that measured at 400 pc.
    
      All these results are based on the gas density as measured from HI data, without correction for the presence of molecular gas (or He), and
the star formation rate as estimated from the \emph{GALEX} UV flux. CO observations are not available for our sample; however, it is known that dwarf galaxies in general have very little CO emission \citep[eg:][]{taylor98}, and that for the usual CO to H$_2$ conversion ratio their ISM is dominated by atomic gas. Given the low ISM metallicity however, it is possible that the standard CO to H$_2$ conversion ratio substantially underestimates the molecular gas in dwarf galaxies \citep[eg:][]{madden97,israel97}, in which case the gas density may change substantially if the molecular gas contribution is considered. It is also likely that the molecular gas fraction increases non linearly with the HI gas column density. As such, it is difficult to make a quantitative prediction of the effect of the adding the contribution of the molecular gas to the total gas density.
Qualitatively however, it is clear that if the gas density is higher than what we estimate, then the discrepancy between the observed \ssfr\  and that predicted by the \cite{ken98} relation increases (i.e. in Figures~\ref{fig:tot},\ref{fig:spc},\ref{fig:plots},\ref{fig:plots200} increasing \sgas\ would move the points to the right, increasing the displacement from the \cite{ken98} relation). An interesting comparison sample in this context is that of \citep{lis02} who measured \ssfr\ and \sgas\ in a sample of tidal dwarfs. The \sgas\ measurements included the contribution of molecular gas, with 8 galaxies having detected molecular gas and 3 galaxies having upper limits on the molecular gas content. The molecular gas fraction was computed using the measured metallicity values. The globally averaged star formation rate that \citet{lis02} compute (from H$\alpha$ observations) shows a similar trend to that seen in Fig.~\ref{fig:tot}, i.e. that most of the galaxies have \ssfr\ lower than predicted by the Kennicutt-Schmidt relation. 

\begin{figure}
\psfig{file=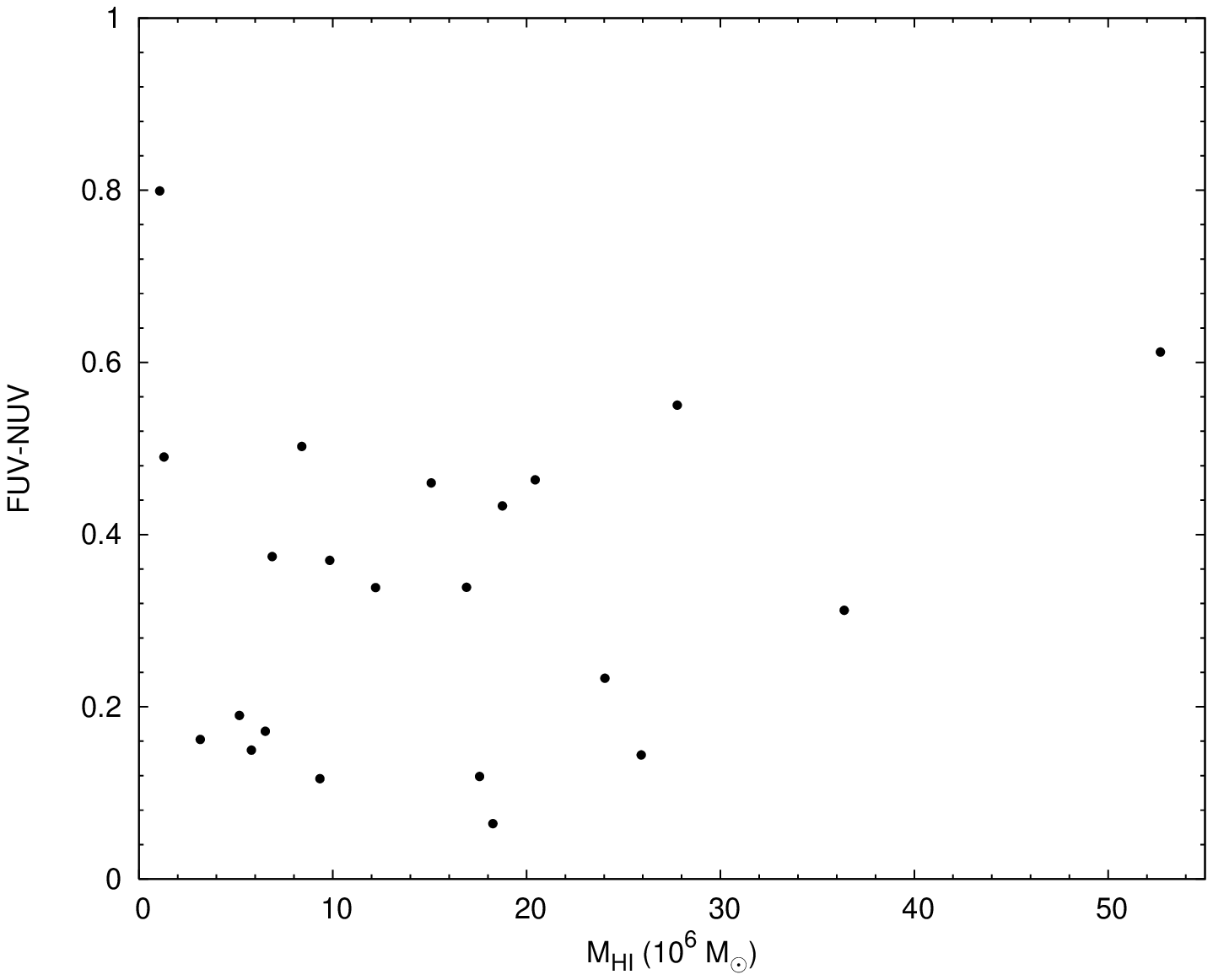,width=3.5truein}
\caption{Disk averaged FUV-NUV color of 22 galaxies in our sample, plotted against the mass of HI gas contained within their optical disks.}
\label{fig:col} 
\end{figure}

     As mentioned in Section~\ref{ssec:uvdata} estimating the SFR from the FUV flux is based on several assumptions, viz. (i)~the stars have solar metallicity and Salpeter IMF, (ii)~the galaxy has had continuous star formation over time scales of 10$^8$ years or longer, and (iii)~internal and external dust attenuation has been correctly
accounted for. For dwarf galaxies as faint as those in our sample, the metallicity is likely to be much lower than solar \citep[eg.][]{lee06}. Lowering the metallicity will lead to a larger UV flux per unit mass of star formation, i.e. our adopted \ssfr\  will be an overestimate of the true \ssfr. Making a quantitative estimate of the magnitude of this underestimate is beyond the scope of this paper, however, we note that the sense is such that it would lead to a further discrepancy between our adopted \ssfr\ and that predicted by the \citep{ken98} relation. Conversely, truncating the IMF at the high mass end would reduce the UV flux per unit mass of star formation, and hence bring the our observations closer to the \citep{ken98} relation. Similarly, if the star formation is dominated by a recent star burst, then the UV flux would be at its peak, and our adopted \ssfr\ would be an over estimate; on the other hand, if the galaxy has not undergone star formation for some time, our adopted calibration would under estimate the \ssfr. Although the detailed star formation history is not available for our sample galaxies, \cite{lee08} shows that for dwarfs (somewhat more massive than our own) the bulk of the star formation does not occur via  bursts. The adopted calibration is also subject to uncertainties in the correction for dust extinction. We have assumed that no correction for internal extinction is necessary. It should be noted that (1) IR fluxes are available for only 7 galaxies in our sample \citep{lis07, wal07, wu06, eng05, vad05} and (2) 
\citet{wal07} have shown that dwarf galaxies have low but non-zero dust content.
Internal dust extinction can in principle be estimated from comparing the IR and UV fluxes \citep[][etc.]{pan03, bua96}, or the FUV to NUV color \citep[][etc.]{boi07, gil07, cor06, sei05}. Infrared fluxes are not available for all our sample galaxies, but NUV data is available for 22 galaxies in our sample (i.e. except for And~IV), and Figure~\ref{fig:col} shows the disk averaged FUV-NUV colors as a function of the HI mass  in the optical disk. Star-forming galaxies with low extinction are expected to have FUV-NUV color close to zero; our sample galaxies however typically have redder colors, with an average FUV-NUV color of 0.34. Redder colors could also be the result of a fading star burst, or a truncated IMF \citep{bos06, cor08}. These issues have been explored in depth by \citet{boi08} who found similarly red UV colours for their sample of low surface brightness galaxies. \citet{boi08} model the variation of FUV-NUV color with time for different  upper cut-offs of mass for the IMF and different metallicities. From their results it appears that either the IMF has to be truncated, or the colors have to be dominated by a fading starburst to reproduce the observed colors, both of which would bring our results closer to the Kennicutt-Schmidt relation. For example, FUV-NUV color of 0.34 would be observed in a galaxy which (1)~has a Kroupa IMF \citep{kro93} with metallicity = Z$_\odot$/20, an age of 6.8$\times$10$^8$ years, and star formation was quenched after 10$^8$ years or (2)~has a Kroupa IMF truncated at 5M$_\odot$ at the highend, solar metallicity, has been forming stars for 4.3$\times$10$^8$ years continuously or (3)~has an age of 2.4$\times$10$^8$ years and star formation was quenched after 10$^8$ years or (4)~Has Kroupa IMF, solar metallicity, an age of 2.4$\times$10$^8$ years and star formation was quenched after 10$^8$ years. For galaxies in our sample, FUV-NUV colours range from $\sim$0.03 to $\sim$0.8, and hence the range of possible models is correspondingly larger.  While we don't explore this issue further, it should be noted that alternate IMFs or star formation histories could lead to revision in the interpretation of our data.

 While there have not been systematic earlier studies of star formation recipes for similarly faint galaxies, it is interesting to compare our results with what is already known for brighter dwarfs or large spiral galaxies. Empirical studies of star formation thresholds have generally used the H$\alpha$ luminosity as a tracer of the current star formation rate. The H$\alpha$ luminosity is sensitive largely to stars with masses $>$~10 M$_\odot$  and lifetimes $<$~20 Myr; one hence gets a measure of the instantaneous star formation. A study of spatially resolved star formation laws for regions of size $\sim$ 500 pc in the spiral galaxy M51a \citep{ken07} which uses both H$\alpha$ and Infrared fluxes to estimate star formation, show very similar results to what we obtain here. There is no evidence of any obvious star formation threshold. 
Interestingly the relation between SFR surface density and HI gas surface density is quite steep, but the power law index comes down to 1.56$\pm$0.04 
after taking molecular gas into
account.
\citet{thi07} did a multi-wavelength study of the spiral galaxy NGC~7331. Along with looking for an azimuthally averaged Schmidt's law, they also show pixel by pixel correlations between \sgas\ and \ssfr\, at different resolutions (400~pc and 1~kpc). Their \ssfr\ vs. \shi\ plot looks strikingly similar to that for our galaxies, with a flat lower portion changing to a steep power law portion, which only reaches the Kennicutt-Schmidt law level at higher column densities. However, unlike the galaxies in our sample, NGC~7331 has a number of pixels lying above the level at which the Kennicutt-Schmidt relation holds. Even after adding molecular gas to \sgas\, the  plot remains qualitatively unchanged, except, as expected, the power law portion becomes less steep and \ssfr\ for a large number of pixels fall below the Kennicutt-Schmidt law level. This is similar to what we expect to find on taking molecular gas into account for our sample. In another recent study, Boissier et al. \citep{boi07} look for evidence for a threshold density for star formation using \emph{GALEX} UV fluxes as a measure of the star formation rate. Similarly to the results presented here, they also do not find any evidence for a threshold below which star formation is completely quenched.
Finally, \citet{wyd09} in a paper published after our manuscript was submitted, find results similar to ours when comparing the global \ssfr\ to \sgas\ for a sample of LSB galaxies.  

\begin{figure}
\psfig{file=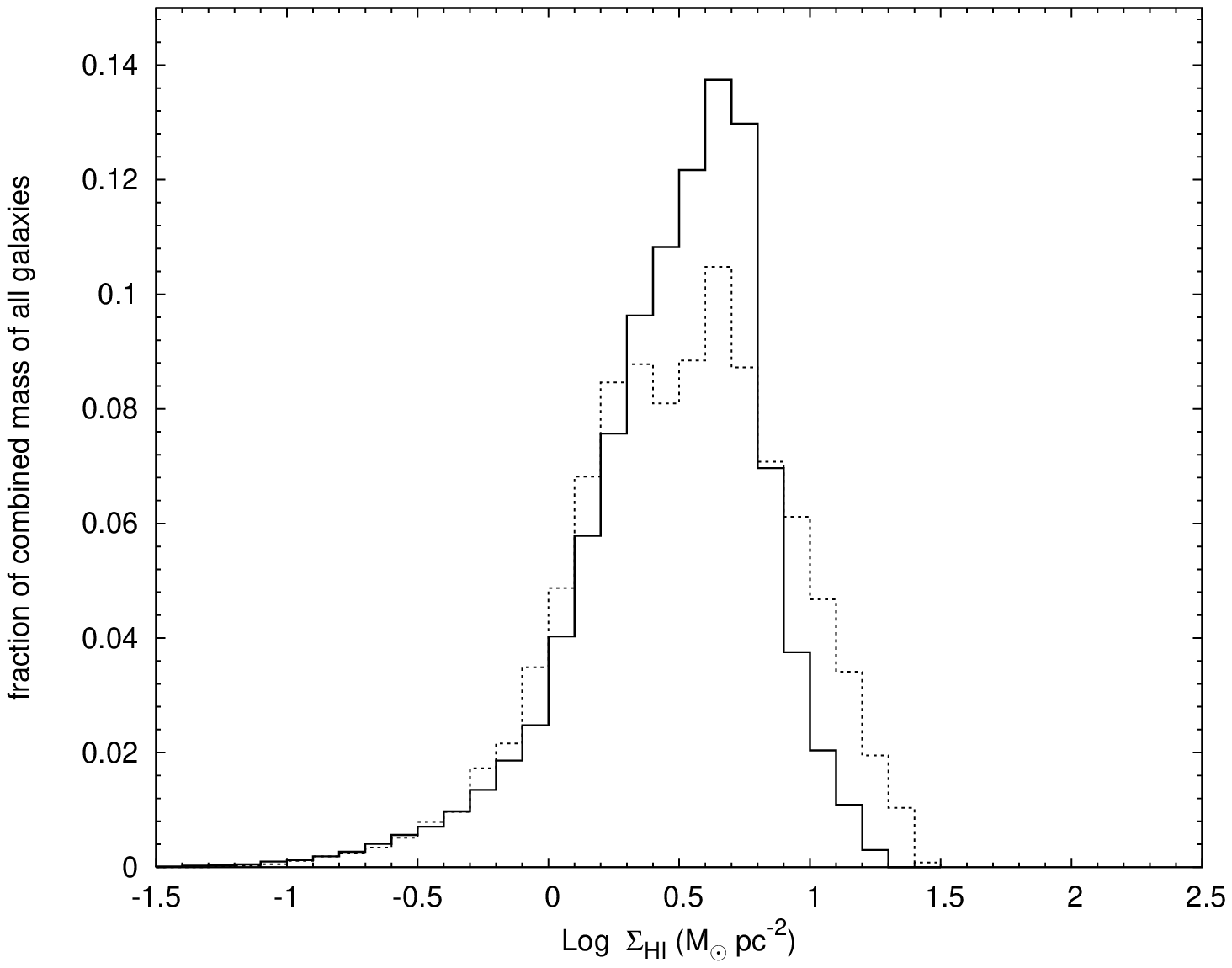,width=3.5truein}
\caption{Histograms depicting what fraction of the combined mass of all galaxies considered lie in a particular column density bin (0.1 dex each). The 
histogram in bold line is for the 22 galaxies (UGC 0685 left out, see text) imaged at linear resolutions of 400 pc. The dashed line histogram is for the sub-sample of 
10 galaxies imaged at linear resolutions of 200 pc.}
\label{fig:hst} 
\end{figure}

   A similar pixel by pixel correlation study (at 220 pc linear resolution) was done for the Magellanic clouds by \cite{ken95}. The star formation rate in that case was traced by H$\alpha$ emission. The (\ssfr,\sgas) scatter plot in \cite{ken95} is qualitatively similar to those in the current paper; the power law indices that they find are $1.75 \pm 0.3$ for the LMC and $1.95 \pm 0.3$ for the SMC. In a recent paper, \citet{bigiel08} study the relation between the star formation rate density and the gas column density in a sample of 18 nearby galaxies at $\sim$ 750 pc resolution using data from the THINGS survey. Their sample covers a range of galaxy types, including dwarfs, but they do not present pixel by pixel correlations for galaxies fainter than the FIGGS magnitude limit of $-$14.9. For the 4 irregular galaxies for which they do fit a power law, they find power law indices that vary from 1.59 to 2.78. They also find that the ``star formation efficiency'' (i.e. \ssfr/\sgas) is lower in dwarfs and the outer parts of spirals than in the inner, H$_2$ dominated regions of spiral galaxies. This is similar to our finding that the SFR in dwarfs is lower than that predicted from the \cite{ken98} relation. They also find that the gas density in dwarfs truncates sharply at about 9M$_\odot$/pc$^2$. We show in Fig.~\ref{fig:hst} the distribution of column densities at both 400 pc and 200 pc resolutions (the data from UGC~685 was not included while calculating the distribution of column densities at 400 pc resolution, as it is unique in having some gas at column densities between $\sim20$ and $\sim30$ M$_\odot$ pc$^{-2}$, and it wasn't imaged at 200 pc resolution). The 400 pc resolution data does indeed show a fall off in the amount of gas at column densities more than $\sim 10$ M$_\odot$/pc$^2$, however at the higher resolution, one can see a tail in the distribution that extends to $\sim 30$ M$_\odot$/pc$^2$. It would appear that dense HI gas occurs in small clumps whose density gets smoothed out when one observes with a coarser resolution.

  \cite{begum06} compared the morphology of the H$\alpha$ emission and the HI emission (at 300 pc resolution) in a sample of 10 similarly faint dwarf galaxies.  Although the H$\alpha$ emission generally occurred near the central regions where the gas column density is high, no one to one relation between the HI column density and the gas column density was found. One possible explanation for this observation is stochasticity in the H$\alpha$ based estimates of \ssfr. There are two reasons why UV based estimates of \ssfr would be less susceptible to stochastic fluctuations than H$\alpha$ based ones, viz. (1)~as mentioned above H$\alpha$ emission traces essentially the instantaneous star formation rate, while UV emission traces the star formation rate over the last $\sim 10^8$ years, and (2)~stochasticity in the formation of high mass stars becomes more important at the low star formation rates that we probe here \citep{oey05}. \citet{pfl08} study a similar phenomena observed in the outer portion of disk galaxies, where UV traces star formation beyond the observed H$\alpha$ cut-off. They show that a local formulation of the concept of clustered star formation naturally leads to a steeper decrease in the H$\alpha$ luminosity as compared to \ssfr\, and hence the H$\alpha$ cut-off arises naturally. In a recent theoretical study \citet{kru08} suggest that massive stars, the likes of which are responsible for H$\alpha$ emission, can only form when the gas column density is 1 g cm$^{-2}$ or higher. But even if the \sgas\ is not so extreme, lower mass stars as traced by UV will continue to form, and hence UV based and not H$\alpha$ based \ssfr\ should be used to look for star formation ''thresholds''.
       
One of our principal findings is that the \ssfr\ that we find are lower than those predicted from relations based on observations of the central regions of galaxies (viz. the ``Kennicutt--Schmidt'' relation).
This suggests that the star formation efficiency is environment dependent, and proceeds differently in dwarf galaxies than in central regions of spirals. At small (i.e. 200 pc) scales offsets between the UV peaks and the HI peaks are more pronounced. Possible reasons for this could be that the gas is predominantly molecular in star forming regions, or that at such small scales the feed back from star formation complicates the relation between \ssfr\ and \sgas. In this context is interesting to note that (1)~the \ssfr\ approaches the value predicted by the \cite{ken98} relation at high gas densities, and (2)~for those galaxies for which a power law parametrization is possible, the power law indices at 200 pc are flatter (i.e. closer to that observed in the centers of spiral galaxies). Note that the average measured gas density is higher in the 200 pc maps than the 400 pc maps (see Fig~\ref{fig:hst}, Table~\ref{tab:res},~\ref{tab:res200}). These facts are consistent with a scenario that when star formation does occur in dense clouds, there is, initially at least, a near universal relation between \ssfr\ and \sgas. Feedback and other effects may lead to a break down of this relation, depending on the exact evolutionary phase at which the observations are made. Even in such a scenario, dwarfs could have a lower average star formation efficiency both because sufficiently dense regions are rarer, and because the effects of feedback are more important.  A proper test of this scenario would require higher angular resolution observations of a diverse sample of star forming galaxies. 

\section*{Acknowledgments}
Some of the data presented in this report were obtained from the Multimission Archive at the Space Telescope Science Institute (MAST). STScI is operated by the Association of Universities for Research in Astronomy, Inc., under NASA contract NAS5-26555. Support for MAST for non-HST data is provided by the NASA Office of Space Science via grant NAG5-7584 and by other grants and contracts.

We thank the staff of the GMRT who have made the observations used in this paper possible.
GMRT is run by the National Centre for Radio Astrophysics of the Tata Institute of Fundamental Research.

\appendix
\section{Pixel by Pixel comparison plots}
\label{sec:ap}

\begin{figure*}
\psfig{file=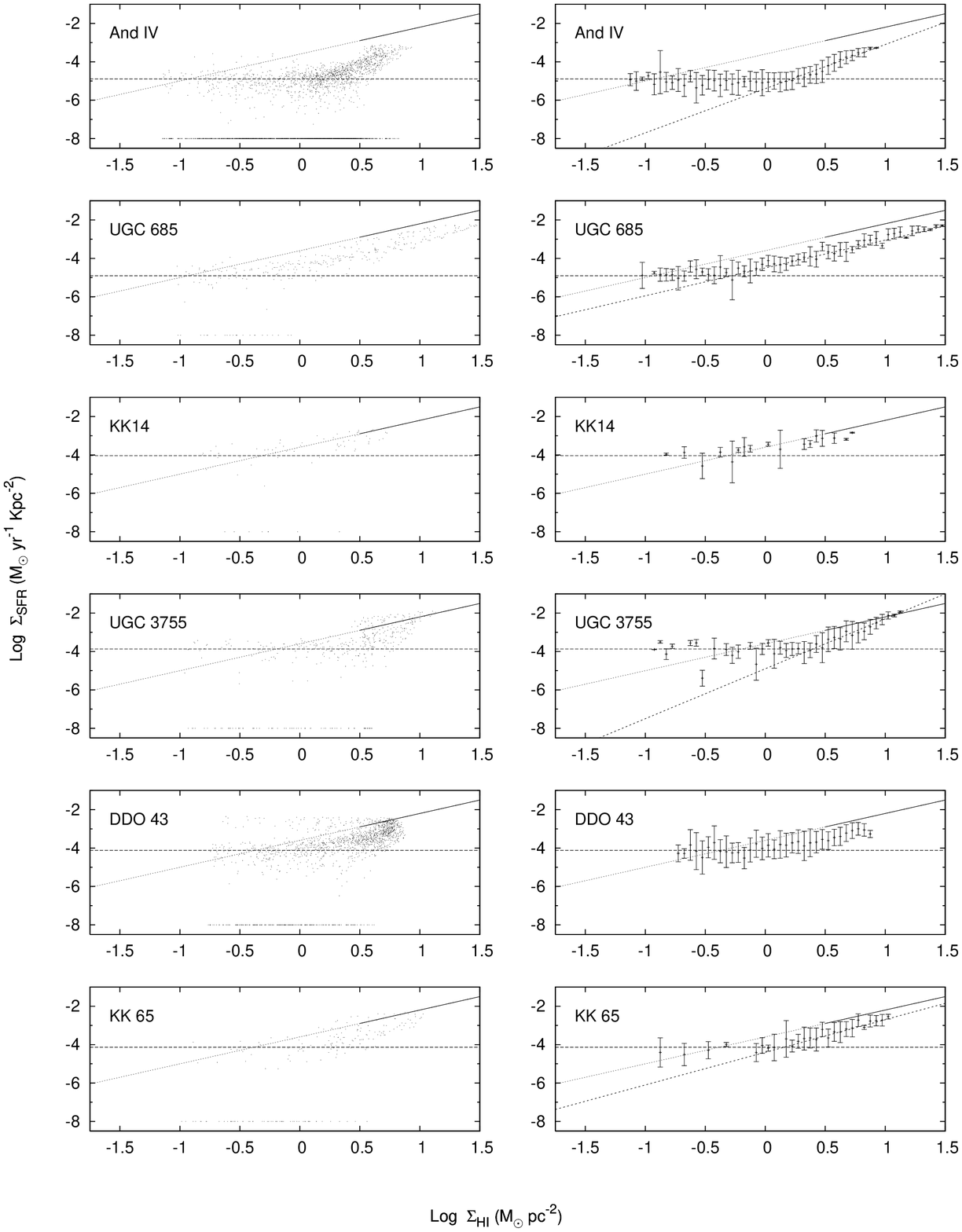,width=6.5truein}
\caption{The pixel by pixel comparison of gas density and SFR  for all galaxies at 400 pc resolution. Each pixel is represented by a point in the left panel, while the right panel shows the binned data. The bin size is 0.05 dex in gas column density. The error bars show the 1$\sigma$ scatter within the bin. The dashed line shows the best fit straight line (see Section~\ref{sec:res}) in all cases in which it was possible to make a meaningful fit. The global ``Kennicutt--Schmidt'' law \citep{ken98} is shown as a solid line above the threshold density for star formation, and as a dotted line below that. The horizontal line marks the ($1\sigma$) sensitivity level of the \emph{GALEX} image. }
\label{fig:plots}
\end{figure*}

\begin{figure*}
\addtocounter{figure}{-1}
\psfig{file=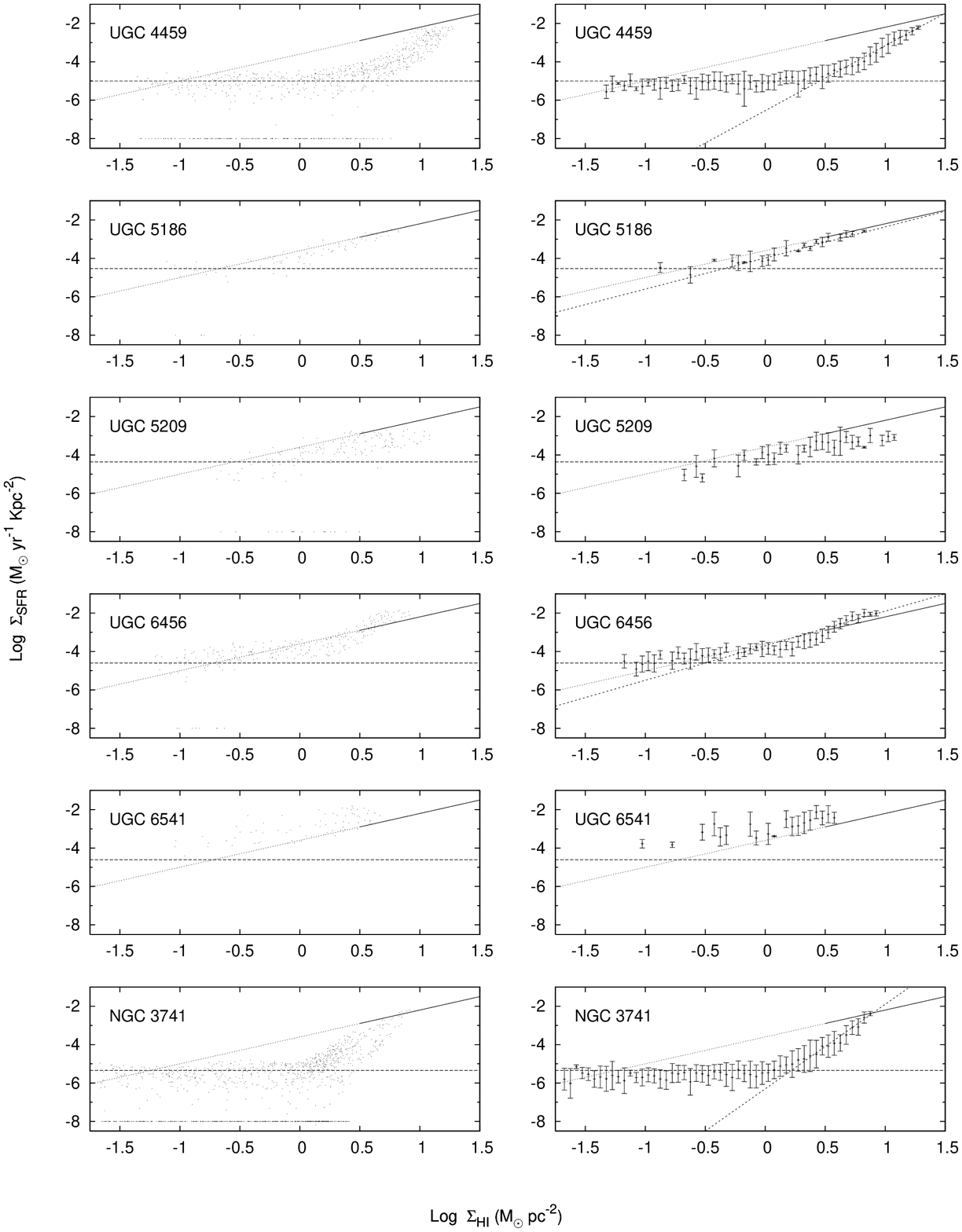,width=6.5truein}
\caption[Continued]{({\it{continued}})}
\label{fig:plots}
\end{figure*}

\begin{figure*}
\addtocounter{figure}{-1}
\psfig{file=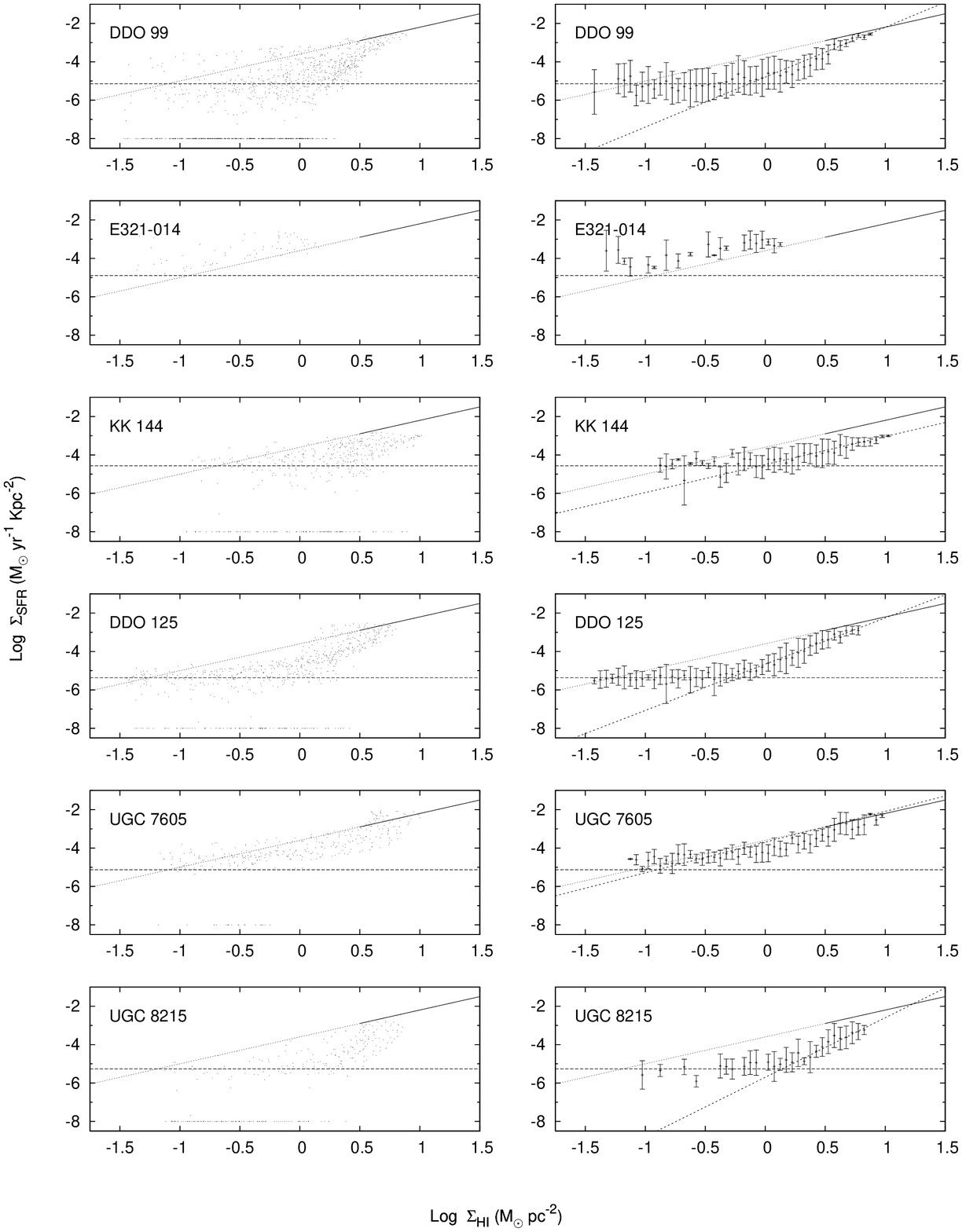,width=6.5truein}
\caption[Continued]{({\it{continued}})}
\label{fig:plots}
\end{figure*}

\begin{figure*}
\addtocounter{figure}{-1}
\psfig{file=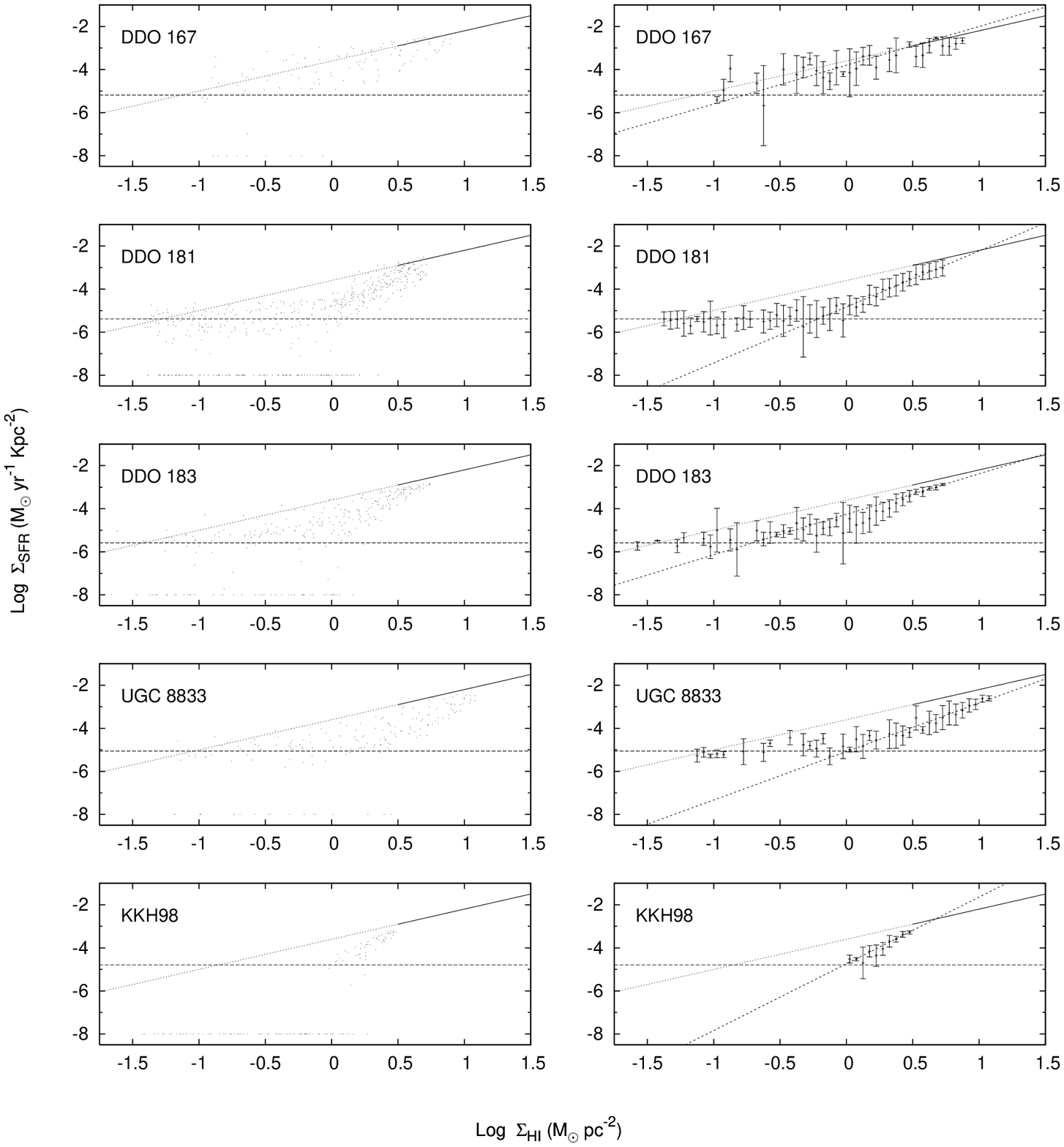,width=6.5truein}
\caption[Continued]{({\it{continued}})}
\label{fig:plots}
\end{figure*}

\begin{figure*}
\psfig{file=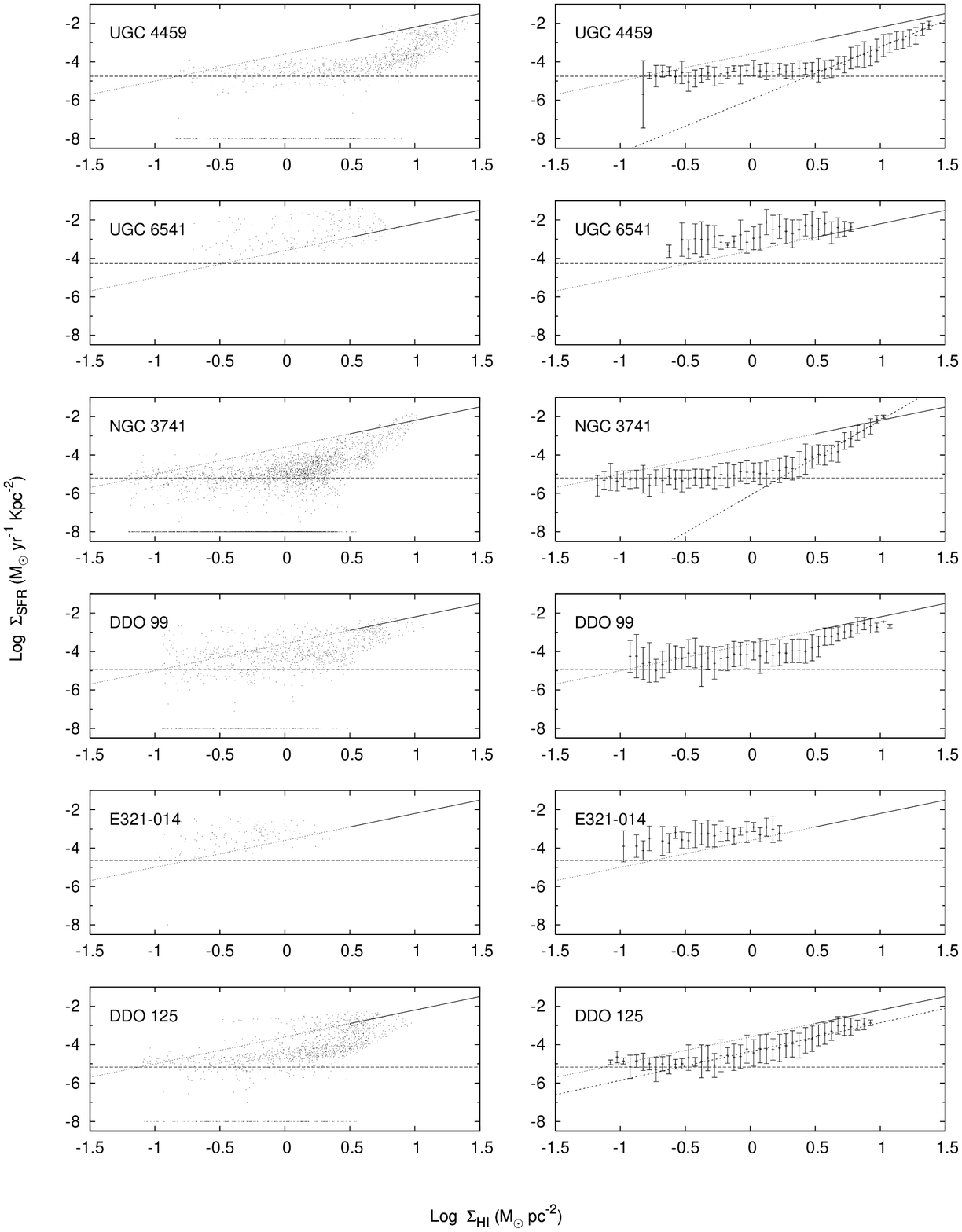,width=6.5truein}
\caption{The pixel by pixel comparison of \sgas\ and \ssfr\  for all galaxies at $\sim$200 pc resolution.. Each pixel is represented by a point in the left panel, while the right panel shows the binned data. The bin size is 0.05 dex in gas column density. The error bars show the 1$\sigma$ scatter within the bin. The dashed line shows the best fit straight line (see Section~\ref{sec:res}) in all cases in which it was possible to make a meaningful fit. The global ``Kennicutt--Schmidt'' law \citep{ken98} is shown as a solid line above the threshold density for star formation, and as a dotted line below that. The horizontal line marks the ($1\sigma$) sensitivity level of the \emph{GALEX} image.}
\label{fig:plots200}
\end{figure*}

\begin{figure*}
\addtocounter{figure}{-1}
\psfig{file=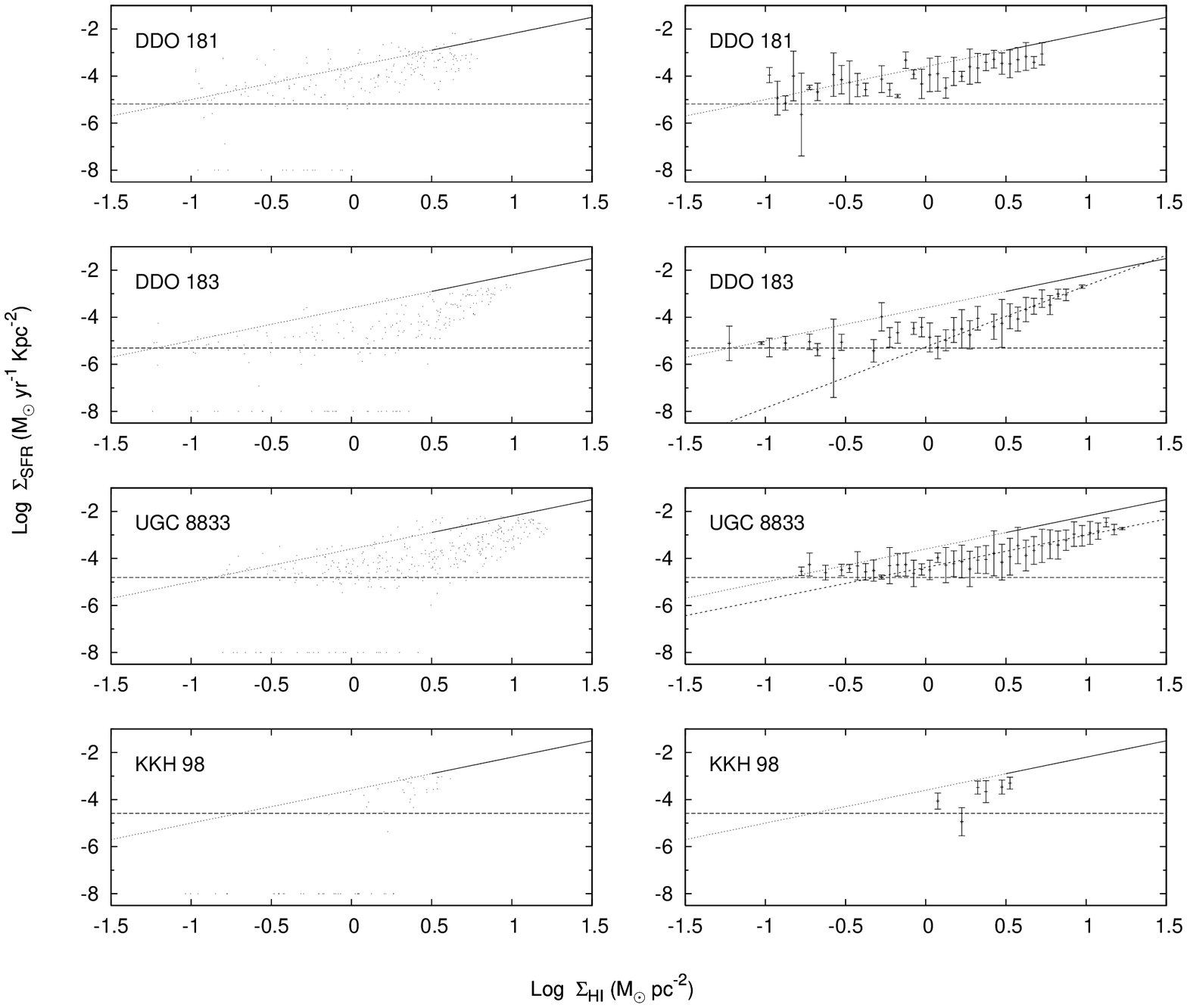,width=6.5truein}
\caption[Continued]{({\it{continued}})}
\label{fig:plots200}
\end{figure*}

\bsp

\label{lastpage}


\begin{thebibliography}{}
\bibitem[\protect\citeauthoryear{Begum \& Chengalur}{2003}]{begum03} Begum A., \& Chengalur J. N., 2003, A\&A, 409, 879
\bibitem[\protect\citeauthoryear{Begum \& Chengalur}{2004}]{begum04} Begum A., \& Chengalur J. N., 2004, BASI, 32, 3, 231
\bibitem[\protect\citeauthoryear{Begum et al.}{2006}]{begum06} Begum A.,
Chengalur J. N., Karachentsev I. D., Kaisin S. S., \& Sharina M. E., 2006, MNRAS, 365, 1220
\bibitem[\protect\citeauthoryear{Begum et al.}{2008}]{ay08} Begum A.,
Chengalur J. N., Karachentsev I. D., Sharina M. E. \& Kaisin S. S., 2008, MNRAS,
386, 1667B
\bibitem[\protect\citeauthoryear{Bigiel et al.}{2008}]{bigiel08} Bigiel F., Leroy A., Walter F., Brinks E., de Blok W. J. G., Madore B., \& Thornley M. D., 2008, AJ, 136, 2846
\bibitem[\protect\citeauthoryear{Boissier et al.}{2007}]{boi07} Boissier S. et al., 2007, ApJS, 173, 524
\bibitem[\protect\citeauthoryear{Boissier et al.}{2008}]{boi08} Boissier S. et al., 2008, ApJ, 681, 244
\bibitem[\protect\citeauthoryear{Boselli et al.}{2006}]{bos06} Boselli A., Boissier S., Cortese L., Gil de Paz A., Seibert M., Madore B. F., Buat V., \& Martin D. C., 2006, ApJ, 651, 811
\bibitem[\protect\citeauthoryear{Buat \& Xu}{1996}]{bua96} Buat V., \& Xu C., 1996, A\&A, 306, 61
\bibitem[\protect\citeauthoryear{Cardelli, Clayton \& Mathis}{1989}]{car89} Cardelli Jason A., Clayton Geoffrey C., \& Mathis John S., 1989, ApJ, 345, 245
\bibitem[\protect\citeauthoryear{Cortese et al.}{2006}]{cor06} Cortese L. et al., 2006, ApJ, 637, 242
\bibitem[\protect\citeauthoryear{Cortese, Gavazzi \& Boselli}{2008}]{cor08} Cortese L., Gavazzi G., \& Boselli A., 2008, MNRAS, 390, 1282
\bibitem[\protect\citeauthoryear{de Blok \& Walter}{2006}]{deB06} de Blok W.J.G., \& Walter F., 2006, AJ, 131, 363
\bibitem[\protect\citeauthoryear{Dong, Lin \& Murray}{2003}]{don03} Dong Shawfeng, Lin D. N. C., \& Murray S. D., 2003, ApJ, 596, 930
\bibitem[\protect\citeauthoryear{Elmegreen \& Hunter}{2006}]{elm06} Elmegreen Bruce G., \& Hunter Deidre A., 2006, ApJ, 636, 712
\bibitem[\protect\citeauthoryear{Engelbracht et al.}{2005}]{eng05} Engelbracht C. W., Gordon K. D., Rieke G. H., Werner M. W., Dale D. A., \& Latter W. B., 2005, ApJ, 628, L29
\bibitem[\protect\citeauthoryear{Gil de Paz et al.}{2007}]{gil07} Gil de Paz Armando et al., 2007, ApJS, 173, 185
\bibitem[\protect\citeauthoryear{Hopp}{1999}]{hop99} Hopp U., 1999, A\&AS, 134, 317
\bibitem[\protect\citeauthoryear{Israel}{1997}]{israel97} Israel F.~P., 1997, A\&A, 328, 471 
\bibitem[\protect\citeauthoryear{Karachentsev et al.}{2004}]{kar04} Karachentsev I. D., Karachentseva V. E., Hutchmeier W. K., \& Makarov D. I., 2004, AJ, 127, 2031
\bibitem[\protect\citeauthoryear{Kennicutt}{1989}]{ken89} Kennicutt Jr. Robert C., 1989, ApJ, 344, 685
\bibitem[\protect\citeauthoryear{Kennicutt}{1997}]{ken97} Kennicutt Jr. Robert C., 1997, in van der Hulst J. M., ed., The Interstellar Medium in Galaxies, Dordrecht: Kluwer, p. 171
\bibitem[\protect\citeauthoryear{Kennicutt}{1998}]{ken98} Kennicutt Jr. Robert C., 1998, ApJ, 498, 541
\bibitem[\protect\citeauthoryear{Kennicutt}{1998a}]{ken98a} Kennicutt Jr. Robert C., 1998a, ARA\&A, 36, 189
\bibitem[\protect\citeauthoryear{Kennicutt et al.}{1995}]{ken95} Kennicutt Jr. Robert C., Bresolin F., Bomans D. J., Bothun G. D., \& Thompson I. B., 1995, AJ, 109, 594
\bibitem[\protect\citeauthoryear{Kennicutt et al.}{2007}]{ken07} Kennicutt Jr. Robert C. et al., 2007, ApJ, 671, 333
\bibitem[\protect\citeauthoryear{Kroupa, Tout \& Gilmore}{1993}]{kro93} Kroupa P., Tout C. A., \& Gilmore G., 1993, MNRAS, 262, 545
\bibitem[\protect\citeauthoryear{Krumholz \& McKee}{2008}]{kru08} Krumholz Mark R., \& McKee Christopher F., Nat, 451, 1082
\bibitem[\protect\citeauthoryear{Krumholz \& Thompson}{2007}]{kru07} Krumholz Mark R., \& Thompson Todd A., 2007, ApJ, 669, 289
\bibitem[\protect\citeauthoryear{Lee et al.}{2006}]{lee06} 
Lee H., Skillman E.~D., Cannon J.~M., Jackson D.~C., Gehrz R.~D., Polomski E.~F., Woodward C.~E., 2006, ApJ, 647, 970 
\bibitem[\protect\citeauthoryear{Lee et al.}{2008}]{lee08} Lee J.~C., Kennicutt R.~C., Jr., Jos{\'e} G.~Funes S.~J., Sakai S., Akiyama S., 2008, preprint (astro-ph/0810.5132)
\bibitem[\protect\citeauthoryear{Li, Mordecai--Mark \& Klessen}{2005}]{li05} Li Yuexing, Mac Low Mordecai--Mark, \& Klessen Ralf S., 2005, ApJ, 620, L19
\bibitem[\protect\citeauthoryear{Lisenfeld et al.}{2002}]{lis02} Lisenfeld Ute, Braine Jonathan, Vallejo Olivier, Duc Pierre-Alain, Leon Stephane, Brinks Elias, \& Charmandaris Vassilis, in Grebel E. K., \& Brander W., eds., ASP Conf. Ser. Vol 285, Modes of Star Formation and the Origin of Field Populations. Astron. Soc. Pac., San Francisco, p. 406 
\bibitem[\protect\citeauthoryear{Lisenfeld et al.}{2007}]{lis07} Lisenfeld U. et al., 2007, A\&A, 462, 507
\bibitem[\protect\citeauthoryear{Madden et al.}{1997}]{madden97} 
Madden S.~C., Poglitsch A., Geis N., Stacey G.~J., Townes C.~H., 1997, ApJ, 
483, 200 
\bibitem[\protect\citeauthoryear{Martin \& Kennicutt}{2001}]{mar01} Martin C.L., \& Kennicutt R.C., 2001, ApJ, 555, 301
\bibitem[\protect\citeauthoryear{Nagamine, Springel \& Hernquist}{2004}]{nag04} Nagamine K., Springel V., \& Hernquist L., 2004, MNRAS, 348, 435
\bibitem[\protect\citeauthoryear{Oey \& Clarke}{2005}]{oey05} Oey M. S., \& Clarke C. J., 2005, ApJ, 620, L43
\bibitem[\protect\citeauthoryear{Panuzzo et al.}{2003}]{pan03} Panuzzo P., Bressan A., Granato G. L., Silva L., \& Danese L., 2003, A\&A, 409, 99 
\bibitem[\protect\citeauthoryear{Pflamm-Altenburg \& Kroupa}{2008}]{pfl08} Pflamm-Altenburg Jan, \& Kroupa Pavel, Nat, 455, 641
\bibitem[\protect\citeauthoryear{Pilyugin}{2001}]{pilyugin01} Pilyugin L.~S., 2001, A\&A, 374, 412 
\bibitem[\protect\citeauthoryear{Quirk}{1972}]{qui72} Quirk W. J., 1972, ApJ, 176 L9
\bibitem[\protect\citeauthoryear{Safronov}{1960}]{safronov60} Safronov V. S., 1960, Annales d'Astrophysique, 23, 979
\bibitem[\protect\citeauthoryear{Schlegel, Finkbeiner \& Davis}{1998}]{sch98} Schlegel David J., Finkbeiner Douglas P., \& Davis Marc, 1998, ApJ, 500, 525
\bibitem[\protect\citeauthoryear{Schmidt}{1959}]{s59} Schmidt M., 1959, ApJ, 129, 243
\bibitem[\protect\citeauthoryear{Seibert et al.}{2005}]{sei05} Seibert Mark et al., 2005, ApJ, 619, L55
\bibitem[\protect\citeauthoryear{Skillman}{1987}]{ski87} Skillman E.D., 1987, in Lonsdale Persson C. J., ed., Star Formation in Galaxies, NASA, p. 263
\bibitem[\protect\citeauthoryear{Spitzer}{1968}]{spi68} Spitzer L., 1968, Diffuse Matter in Space, New York: Wiley
\bibitem[\protect\citeauthoryear{Springel}{2000}]{spr00} Springel Volker, 2000, MNRAS, 312, 859
\bibitem[\protect\citeauthoryear{Springel et al.}{2005}]{spr05} Springel Volker et al., 2005, Nat, 435, 629
\bibitem[\protect\citeauthoryear{Taylor, Kobulnicky, \& Skillman}{1998}]{taylor98} Taylor C.~L., Kobulnicky H.~A., Skillman E.~D., 1998, AJ, 116, 2746
 \bibitem[\protect\citeauthoryear{Thilker et al.}{2007}]{thi07} Thilker David A. et al., 2007, ApJ, 173, 572
\bibitem[\protect\citeauthoryear{Toomre}{1964}]{t64} Toomre A., 1964, ApJ, 139, 1217
\bibitem[\protect\citeauthoryear{Vaduvescu et al.}{2005}]{vad05} Vaduvescu Ovidiu, McCall Marshall L., Richer Michael G., \& Fingerhut Robin L., 2005, AJ, 130, 1593
\bibitem[\protect\citeauthoryear{Walter et al.}{2007}]{wal07} Walter Fabian et al., 2007, ApJ, 661, 102
\bibitem[\protect\citeauthoryear{Wyder et al.}{2009}]{wyd09} Wyder Ted K. et al., 2009, arXiv:0903.3015
\bibitem[\protect\citeauthoryear{Wu et al.}{2006}]{wu06} Wu Yanling, Charmandaris V., Hao Lei, Brandl B. R., Bernard--Salas J., Spoon H. W. W., \& Houck J. R., 2006, ApJ, 639, 157
\bibitem[\protect\citeauthoryear{Young et al.}{2003}]{young03} Young L. M., van Zee L., Lo K. Y., Dohm--Palmer R. C., \& Beierle Michelle E., 2003, ApJ, 592, 111
\end{thebibliography}
\end{document}